\documentclass[pra,twocolumn,preprintnumbers,amsmath,amssymb]{revtex4}
\usepackage{graphics}
\usepackage{psfrag}
\usepackage{hyperref}

\begin{document}

\title{Functional renormalization for Bose-Einstein Condensation}
%\date{\today}
\author{S. Floerchinger}
\author{C. Wetterich}
\affiliation{Institut f\"{u}r Theoretische Physik\\Universit\"at Heidelberg\\Philosophenweg 16, D-69120 Heidelberg}

\begin{abstract}
We investigate Bose-Einstein condensation for interacting bosons at zero and nonzero temperature. Functional renormalization provides us with a consistent method to compute the effect of fluctuations beyond the Bogoliubov approximation. For three-dimensional dilute gases, we find an upper bound on the scattering length $a$ which is of the order of the microphysical scale - typically the range of the Van der Waals interaction. In contrast to fermions near the unitary bound, no strong interactions occur for bosons with approximately pointlike interactions, thus explaining the high quantitative reliability of perturbation theory for most quantities. For zero temperature we compute the quantum phase diagram for bosonic quasiparticles with a general dispersion relation, corresponding to an inverse microphysical propagator with terms linear and quadratic in the frequency. We compute the temperature dependence of the condensate and particle density $n$, and find for the critical temperature $T_c$ a deviation from the free theory, $\Delta T_c/T_c=2.1 a n^{1/3}$. For the sound velocity at zero temperature we find very good agreement with the Bogoliubov result, such that it may be used to determine the particle density accurately.
\end{abstract}


\maketitle

%\tableofcontents

%\newpage
\section{Introduction}
For ultracold dilute non-relativistic bosons in three dimensions, Bogoliubov theory gives a successful description of most quantities of interest \cite{Bogoliubov}. This approximation breaks down, however, near the critical temperature for the phase transition, as well as for the low temperature phase in lower dimensional systems, due to the importance of fluctuations. One would therefore like to have a systematic extension beyond the Bogoliubov theory, which includes the fluctuation effects beyond the lowest order in a perturbative expansion in the scattering length. Such extensions have encountered obstacles in the form of infrared divergences in various expansions \cite{BeliaevGavoretNepomnyashchii}. Only recently, a satisfactory framework has been found to cure these problems \cite{Castellani, Wetterich:2007ba}.

Functional renormalization \cite{FunctionalRenormalization} for the average action \cite{Wetterich:1990an,Wetterich:1992yh,Berges:2000ew} systematically copes with the infrared problems by exploring the difficult long range behavior gradually by means of non-perturbative flow equations, which are based on an exact renormalization group equation \cite{Wetterich:1992yh}. For zero temperature they lead to a consistent description of Bose-Einstein condensation and the quantum phase transition for nonrelativistic bosons in arbitrary dimension \cite{Wetterich:2007ba,Dupuis}. A key ingredient is the generation of a term in the inverse propagator that is quadratic in the frequency $\sim \tilde{v}\omega^2$. It characterizes the long time behavior of the full propagator at long distances, even if no quadratic frequency dependence is present in the microphysical or classical propagator. The coupling $\tilde{v}$ is due to quantum fluctuations and appears during the flow from microphysics to macrophysics. The extreme infrared limit is governed by a "relativistic" model where $\tilde{v}$ dominates, with an enhanced (approximate) space-time symmetry $SO(1,d)$ corresponding to Lorentz symmetry. In one or two dimensions the relativistic term $\sim \tilde{v}$ in the inverse propagator is crucial for a correct description of the low temperature behavior. In contrast, in three dimensions the flow towards the extreme infrared behavior is only logarithmic, such that the quantitative influence remains quite moderate for practical purposes. Nevertheless, the effective coupling $\tilde{v}$ is needed in order to avoid the infrared problems.

In this paper, we extend this formalism to a nonvanishing temperature. We present a quantitative rather accurate picture of Bose-Einstein condensation in three dimensions and find that the Bogoliubov approximation is indeed valid for most quantities. The same method can be applied for one or two dimensions, such that the present work can also serve as a test of the method. This is nontrivial, because insufficient truncations of the flow equations can lead to fake dependencies on the microscopic physics.

For dilute non-relativistic bosons in three dimensions we find an upper bound on the scattering length $a$. This is similar to the "triviality bound" for the Higgs scalar in the standard model of elementary particle physics. As a consequence, the scattering length is at most of the order of the inverse effective ultraviolet cutoff $\Lambda^{-1}$, which indicates the breakdown of the pointlike approximation for the interaction at short distances. Typically, $\Lambda^{-1}$ is of the order of the range of the Van der Waals interaction. For dilute gases, where the interparticle distance $n^{-1/3}$ is much larger than $\Lambda^{-1}$, we therefore always find a small concentration $c=a n^{1/3}$. This provides for a small dimensionless parameter, and perturbation theory in $c$ becomes rather accurate for most quantities. For typical experiments with ultracold bosonic alkali atoms one has $\Lambda^{-1}\approx 10^{-7} \,\text{cm}$, $n^{1/3}\approx 10^4 \,\text{cm}^{-1}$, such that $c\lesssim 10^{-3}$ is really quite small.

Bosons with pointlike interactions can also be employed for an effective description of many quantum phase transitions at zero temperature, or phase transitions at low temperature $T$. In this case, they correspond to quasi-particles, and their dispersion relation may differ from the one of non-relativistic bosons, $\omega=\frac{\vec{p}^2}{2M}$. We describe the quantum phase transitions for a general microscopic dispersion relation, where the inverse classical propagator in momentum and frequency space takes the form $G_0^{-1}=-S\omega-V\omega^2+\vec{p}^2$ (in units where the particle mass $M$ is set to $1/2$). We present the quantum phase diagram at $T=0$ in dependence on the scattering length $a$ and a dimensionless parameter $\tilde{v}\sim V/S^2$, which measures the relative strength of the term quadratic in $\omega$ in $G_0^{-1}$. In the limit $S\rightarrow 0$ ($\tilde{v}\rightarrow\infty$) our model describes relativistic bosons.

A nonzero temperature $T$ introduces a new scale into the problem. This modifies the effective scaling behavior. In particular, near the critical temperature $T_c$, where the Bose-Einstein condensate dissolves, the long distance physics can be described by classical statistics. The non-relativistic bosons belong to the universality class of the three-dimensional $O(2)$ model, with the associated universal critical exponents. The flow equations in the classical regime are well studied and lead, for an appropriate level of the truncation, to very accurate results for the critical behavior \cite{Berges:2000ew,CE}. Here we use a rather simple truncation, but we obtain nevertheless a reasonable description of the non-analytical critical behavior. However, the temperature range for the applicability of the universal critical behavior is found to be very small for the small values of $c$ encountered for the dilute gases.

The value of the critical temperature for interacting bosons cannot be computed within the Bogoliubov theory. We find a temperature shift $\Delta T_c/T_c=2.1 a n^{1/3}$, compared to the free theory. In contrast to other estimates, which only take the classical fluctuations into account (only the zero Matsubara frequency), we include the full quantum statistics (all Matsubara frequencies). We finally compute the sound velocity at $T=0$. This quantity agrees to high precision with the Bogoliubov result. For a known scattering length $a$, it can therefore be used as a precise measure of the density.

Our paper is organized as follows. As a starting point, we specify our microscopic model in Sec. \ref{sectMicroscopicmodel}. In the following Sec. \ref{sectNonperturbativeflowequations}, we recall the method of functional renormalization, explain our truncation of the effective average action and also describe the projection of the exact flow equation onto that truncation. The flow equations in the vacuum are investigated in Sec. \ref{sectUpperboundforthescatteringlength} and an upper bound for the scattering length is derived. In Sec. \ref{sectQuantumphasediagram} we explain our method to determine the density and also study the quantum phase diagram at zero temperature. The effects of a non-vanishing temperature on the condensate are taken into account in Sec. \ref{sectTemperaturedependenceofcondensate}, while Sec. \ref{sectCriticaltemperature} gives our result for the critical temperature in dependence of the interaction strength. Finally, we investigate the sound velocity and draw conclusions in Secs. \ref{sectSoundvelocity} and \ref{sectConclusions}. 

The Appendices \ref{sectTruncation} and \ref{sectSymmetriesandNoethercurrents} contain a motivation of our truncation in terms of a systematic derivative expansion and an analysis of symmetries constraining the form of the effective action. Our explicit results for the flow equations of the effective potential and the kinetic coefficients are shown in Appendices \ref{secFlowEffectivePotential} and \ref{sectFlowofkineticcoefficients}. Appendix \ref{sectPropDisp} analyzes the dispersion relation.

\section{Microscopic model}
\label{sectMicroscopicmodel}
Our microscopic action describes nonrelativistic bosons, with an effective interaction between two particles given by a contact potential. It is assumed to be valid on length scales where the microscopic details of the interaction are irrelevant and the scattering length  is sufficient to characterize the interaction. The microscopic action reads
\begin{equation}
S[\phi]=\int_x \,{\Big \{}\phi^*\,(S\partial_\tau-V\partial_\tau^2-\Delta-\sigma)\,\phi\,+\,\frac{1}{2}\lambda(\phi^*\phi)^2{\Big \}},
\label{microscopicaction}
\end{equation}
with 
\begin{equation}
x=(\tau,\vec{x}), \,\,\int_x=\int_0^{\frac{1}{T}}d\tau\int d^3x.
\end{equation}
The integration goes over the whole space as well as over the imaginary time $\tau$, which at finite temperature is integrated on a circle of circumference $\beta=1/T$ according to the Matsubara formalism. We use natural units $\hbar=k_B=1$. We also scale time and energy units with appropriate powers of $2M$, with $M$ the particle mass. In other words, our time units are set such that effectively $2M=1$. In particular, we use the rescaled chemical potential $\sigma=2M\mu$ and $T$ stands for the temperature multiplied by $2M$. In these units time has the dimension of length squared. For standard non-relativistic bosons one has $V=0$ and $S=1$, but we also consider quasiparticles with a more general dispersion relation described by nonzero $V$.

After Fourier transformation, the kinetic term reads
\begin{equation}
\int_q \phi^*(q)(i S q_0+V q_0^2+\vec{q}^2)\phi(q),
\label{eqMicroscopicFourier}
\end{equation}
with
\begin{eqnarray}
q=(q_0,\vec{q}),\quad \int_q=\int_{q_0}\int_{\vec{q}},\quad \int_{\vec{q}}=\frac{1}{(2\pi)^3}\int d^3q.
\end{eqnarray}
At nonzero temperature, the frequency $q_0=\omega_n=2\pi T n$ is discrete, with
\begin{equation}
\int_{q_0}=T \sum_{n=-\infty}^\infty,
\end{equation}
while at zero temperature this becomes
\begin{equation}
\int_{q_0}=\frac{1}{2\pi}\int_{-\infty}^\infty dq_0.
\end{equation}
The dispersion relation encoded in eq. \eqref{eqMicroscopicFourier} obtains by analytic continuation
\begin{equation}
S\omega+V\omega^2=\vec{q}^2/2M.
\end{equation}

In this paper, we consider homogeneous situations, i.e. an infinitely large volume without a trapping potential. Many of our results can be translated to the inhomogeneous case in the framework of the local density approximation. One assumes that the length scale relevant for the quantum and statistical fluctuations is much smaller than the characteristic length scale of the trap. In this case, our results can be transferred by taking the chemical potential position dependent in the form $\sigma\left(\vec{x})=2M(\mu-V_t(\vec{x}\right))$, where $V_t(\vec{x})$ is the trapping potential.

The microscopic action \eqref{microscopicaction} is invariant under the global $U(1)$ symmetry which is associated to the conserved particle number,
\begin{equation}
\phi\rightarrow e^{i\alpha}\phi.
\end{equation}
On the classical level, this symmetry is broken spontaneously when the chemical potential $\sigma$ is positive. In this case, the minimum of $-\sigma\phi^*\phi+\frac{1}{2}\lambda(\phi^*\phi)^2$ is situated at $\phi^*\phi=\frac{\sigma}{\lambda}$. The ground state of the system is then characterized by a macroscopic field $\phi_0$, with $\phi_0^*\phi_0=\rho_0=\frac{\sigma}{\lambda}$. It singles out a direction in the complex plane and thus breaks the $U(1)$ symmetry. Nevertheless, the action itself and all modifications due to quantum and statistical fluctuations respect the symmetry. For $V=0$ and $S=1$, the situation is similar for Galilean invariance. At zero temperature, we can perform an analytic continuation to real time and the microscopic action \eqref{microscopicaction} is then invariant under transformations that correspond to a change of the reference frame in the sense of a Galilean boost. It is easy to see that in the phase with spontaneous $U(1)$ symmetry breaking also the Galilean symmetry is broken spontaneously: A condensate wave function, that is homogeneous in space and time, would be represented in momentum space by
\begin{equation} 
\phi(\omega,\vec{p})=\phi_0 \,(2\pi)^4\, \delta^{(3)}(\vec{p})\delta(\omega).
\end{equation}
Under a Galilean boost transformation with a boost velocity $2\vec{q}$, this would transform according to
\begin{eqnarray}
\nonumber
\phi(\omega,\vec{p})\rightarrow&&\phi(\omega-\vec{q}^2,\vec{p}-\vec{q})\\
&&=\phi_0\,(2\pi)^4\,\delta^{(3)}(\vec{p}-\vec{q})\delta(\omega-\vec{q}^2).
\end{eqnarray} 
This shows that the ground state is not invariant under such a change of reference frame. This situation is in contrast to the case of a relativistic Bose-Einstein condensate, like the Higgs boson field after electroweak symmetry breaking. A relativistic scalar transforms under Lorentz boost transformations according to
\begin{equation}
\phi(p^\mu)\rightarrow\phi((\Lambda^{-1})^\mu_{\,\,\nu}\,p^\nu),
\end{equation}
such that a condensate wave function
\begin{eqnarray}
\nonumber
\phi_0\,(2\pi)^4 \,\delta^{(4)}(p^\mu)\rightarrow&&\phi_0\,(2\pi)^4\,\delta^{(4)}((\Lambda^{-1})^\mu_{\,\,\nu}\,p^\nu)\\
&&=\phi_0\,(2\pi)^4\, \delta^{(4)}(p^\mu)
\end{eqnarray}
transforms into itself. We will investigate the implications of Galilean symmetry for the form of the effective action in app. \ref{sectTruncation}. An analysis of general coordinate invariance in nonrelativistic field theory can be found in \cite{SonWingate}.

\section{Non-perturbative flow equations}
\label{sectNonperturbativeflowequations}

\subsection{Functional Renormalization Group and Flow equation}
We start with a functional integral representation of the grand canonical partition function
\begin{equation}
Z=\text{Tr}\, e^{-\beta(H-\mu N)}=\int D\chi\, e^{-S[\chi]}.
\label{eqpartfunction}
\end{equation}
In this paper, we work with the formalism of quantum statistics for many particle problems. In contrast to classical statistics, the fields $\chi(\tau,\vec{x})$ are parametrized by an Euclidean time variable $\tau$ in addition to the space variable $\vec{x}$. This Euclidean time is wrapped up on a circle of circumference $\beta=\frac{1}{T}$, such that the fields $\chi(\tau,\vec{x})$ live on a (generalized) torus. The microscopic action $S[\chi]$ is of a form similar to \eqref{microscopicaction}.

We generalize eq. \eqref{eqpartfunction} by introducing a source $J$ for the fields $\chi$ and write
\begin{equation}
Z[J]=e^{W[J]}=\int D \chi \,e^{-S[\chi]+\int_x J\chi}.
\label{eqpartfunctwithJ}
\end{equation}
The $n$-point correlation functions can now be obtained by functional differentiation of $Z[J]$, while $W[J]$ generates the connected $n$-point functions. For example, the expectation value of $\chi$ obtains from
\begin{equation}
\phi(x)=\langle \chi(x)\rangle=\frac{\delta W[J]}{\delta J(x)}.
\end{equation}
The thermodynamic potential $\Phi_G$ associated with the grand canonical partition function is given by
\begin{equation}
\Phi_G=-\frac{1}{\beta}\text{ln}\,Z=-\frac{1}{\beta}\, W[J=0].
\end{equation}

The effective action $\Gamma[\phi]$ is defined as
\begin{equation}
\Gamma[\phi]=\left(-W[J]+\int_x J \phi\right)_{J=J_{\text{ex}}[\phi]}.
\end{equation}
where $J_{\text{ex}}$ is obtained by the inversion of 
\begin{equation}
\frac{\delta W[J]}{\delta  J}\bigg{|}_{J=J_\text{ex}}=\phi.
\end{equation}
It is straightforward to show $\frac{\delta \Gamma[\phi]}{\delta \phi}=J$. In the absence of a source $J$, we obtain for the thermodynamic potential of the grand canonical partition function
\begin{equation}
\beta \Phi_G =\Gamma_{\text{min}}=\Gamma[\phi_{\text{eq}}]
\end{equation}
where $\phi_{\text{eq}}$ is defined by 
\begin{equation}
\frac{\delta \Gamma[\phi]}{\delta \phi}\bigg{|}_{\phi=\phi_{\text{eq}}}=0.
\end{equation}
The effective action $\Gamma[\phi]$ is the generating functional of the one-point irreducible correlation functions and in a sense, its precise knowledge corresponds to the solution of the theory. 

Our method determines $\Gamma[\phi]$ with the help of an exact flow equation. For that purpose, we include an infrared cutoff term $\Delta S_k[\chi]$ in eq. \eqref{eqpartfunctwithJ} and define
\begin{equation}
e^{W_k[J]}=\int D\chi\, e^{-S[\chi]-\Delta S_k[\chi]+\int J \chi}.
\end{equation}
In Fourier space, the cutoff term reads
\begin{equation}
\Delta S_k[\chi]=\int_q R_k(q) \chi^*(q)\chi(q)
\end{equation}
and has the properties
\begin{eqnarray}
\nonumber
R_k(q) & \rightarrow\infty & \quad(k\rightarrow\infty),\\
\nonumber
R_k(q) & \approx k^2 & \quad(q\rightarrow0),\\
R_k(q) & \rightarrow0 & \quad(k\rightarrow0).
\end{eqnarray}
The effective average action is defined as a modified Legendre transform of $W_k$
\begin{equation}
\Gamma_k[\phi]=\left( -W_k[J]+\int_x J\phi \right)_{J=J_\text{ex}}-\Delta S_k[\phi].
\end{equation}
It has the important property, that it interpolates between the microscopic action $S[\phi]$ and the full effective action $\Gamma[\phi]$
\begin{eqnarray}
\nonumber
\Gamma_{k}[\phi] & \rightarrow S[\phi] & \quad (k\rightarrow \infty),\\
\Gamma_{k}[\phi] & \rightarrow \Gamma[\phi] & \quad (k\rightarrow 0).
\end{eqnarray}
In physical terms $\Gamma_k[\phi]$ is the effective action in the presence of an infrared cutoff at a momentum scale $k$. Only fluctuations with momenta larger than $k$ are included. For example, a finite volume $V\sim \frac{1}{k^3}$ of the system under consideration would lead to an situation that is described by $\Gamma_k[\phi]$. 

Our method to determine $\Gamma_k[\phi]$ (and for $k\rightarrow 0$ also $\Gamma[\phi]$) relies on the existence of an exact flow equation \cite{Wetterich:1992yh, Berges:2000ew}
\begin{equation}
\partial_k\,\Gamma_k=\frac{1}{2}\text{Tr}(\Gamma_k^{(2)}+R_k)^{-1}\partial_k R_k.
\label{eqFlowequation}
\end{equation}
Here the trace operation includes a momentum integration $\int_q$, as well as a sum over internal indices $i=1,2$, according to the two real components in the decomposition $\phi(x)=\frac{1}{\sqrt{2}}(\phi_1(x)+i\phi_2(x))$. On the r.h.s. of \eqref{eqFlowequation}, $\Gamma_k^{(2)}$ is the second functional derivative of $\Gamma_k[\phi]$
\begin{equation}
(\Gamma_k^{(2)}[\phi])_{ij}(q,p)=\frac{\overset{\rightharpoonup}{\delta}}{\delta\phi_i(-q)}\Gamma_k[\phi]\frac{\overset{\leftharpoonup}{\delta}}{\delta \phi_j(p)}.
\end{equation}
It is therefore a matrix in internal and momentum space. Correspondingly, $R_k$ in eq. \eqref{eqFlowequation} stands for $R_k(q)\delta_{ij}\delta(q-p).$
The flow equation \eqref{eqFlowequation} describes the evolution of the effective average action with the cutoff scale $k$. 

The functional differential equation \eqref{eqFlowequation} is hard to be solved exactly. In principle, $\Gamma_k$ is described by infinitely many couplings. We will use here an approximation with only a finite number of couplings. This is achieved by an ansatz for a specific form of $\Gamma_k[\phi]$. In the absence of anomalies the effective action $\Gamma[\phi]$ is invariant under the same symmetries as the microscopic action $S[\phi]$. This holds also for the effective average action $\Gamma_k[\phi]$, provided that the cutoff term $\Delta S_k[\phi]$ is also invariant. Our ansatz will respect all symmetries of the classical action.

\subsection{Truncation}
Approximate solutions of the exact flow equations obtain from a truncation of the general form of the effective action. We use here terms with up to two derivatives and truncate
\begin{eqnarray}
\nonumber
\Gamma_k&=&\int_x\bigg{\{} \bar{\phi}^*\left(\bar{S}\partial_\tau-\bar{A}\Delta-\bar{V}\partial_\tau^2\right)\bar{\phi}\\
&&+2\bar{V}(\sigma-\sigma_0)\,\bar{\phi}^*\left(\partial_\tau-\Delta\right)\bar{\phi}+\bar{U}(\bar{\rho},\sigma)\bigg{\}},
\end{eqnarray}
with $\bar{\rho}=\bar{\phi}^*\bar{\phi}$. This particular form is motivated by a more systematic derivative expansion and an analysis of symmetry constraints (Ward identities) in appendix \ref{sectTruncation}. We introduce the renormalized fields $\phi=\bar{A}^{1/2}\bar{\phi}$, $\rho=\bar{A}\bar{\rho}$, the renormalized kinetic coefficients $S=\frac{\bar{S}}{\bar{A}}$, $V=\frac{\bar{V}}{\bar{A}}$ and we express the effective potential in terms of the renormalized invariant $\rho$, with
\begin{equation}
U(\rho,\sigma)=\bar{U}(\bar{\rho},\sigma).
\end{equation}
This yields
\begin{eqnarray}
\nonumber
\Gamma_k &=& \int_x\bigg{\{} \phi^*\left(S\partial_\tau-\Delta-V\partial_\tau^2\right)\phi\\
&&+2V(\sigma-\sigma_0)\, \phi^*\left(\partial_\tau-\Delta\right)\phi+U(\rho,\sigma)\bigg{\}}.
\label{eqSimpleTruncation}
\end{eqnarray}

For the effective potential, we use an expansion around the $k$-dependent minimum $\rho_0(k)$ of the effective potential and the $k$-independent value of the chemical potential $\sigma_0$ that corresponds to the physical particle number density $n$. We determine $\rho_0(k)$ and $\sigma_0$ by the requirements
\begin{eqnarray}
\nonumber
(\partial_\rho U)(\rho_0(k),\sigma_0)&=0\quad&\text{for all}\,k\\
-(\partial_\sigma U)(\rho_0,\sigma_0)&=n\quad&\text{at}\,k=0.
\end{eqnarray}
More explicitly we take a truncation for $U(\rho,\sigma)$ of the form
\begin{eqnarray}
\nonumber
U(\rho,\sigma)&=&U(\rho_0,\sigma_0)-n_k(\sigma-\sigma_0)\\
\nonumber
&&+\left(m^2+\alpha(\sigma-\sigma_0)\right)(\rho-\rho_0)\\
&&+\frac{1}{2}\left(\lambda+\beta(\sigma-\sigma_0)\right)(\rho-\rho_0)^2.
\end{eqnarray}
In the symmetric phase we have $\rho_0=0$, while in the phase with spontaneous symmetry breaking, we have $m^2=0$.
In summary, the flow of $\Gamma_k$ for fixed $\sigma=\sigma_0$ is described by four running renormalized couplings $\rho_0$, $\lambda$, $S$ and $V$. In addition, we need the anomalous dimension $\eta=-k\,\partial_k \text{ln}\bar{A}$. A computation of $n$ requires a flow equation of $n_k$, which involves the couplings linear in $\sigma-\sigma_0$, namely $\alpha$ and $\beta$. The pressure is calculated by following the $k$-dependence of the height of the minimum $p_k=-U(\rho_0,\sigma_0)$. All couplings $\rho_0$, $\lambda$, $S$, $V$, $\bar{A}$, $n_k$, $p_k$, $\alpha$, $\beta$ depend on $k$ and $T$. The physical renormalized couplings obtain for $k\rightarrow 0$. They specify the thermodynamic potential $U(\rho_0,\sigma_0)$ as well as suitable derivatives of the potential and the correlation function. 

The "initial values" at the scale $k=\Lambda$ are determined by the requirement
\begin{equation}
\Gamma_\Lambda[\phi]=S[\phi],
\end{equation}
using the microscopic action $S[\phi]$ in eq. \eqref{microscopicaction}. This implies the initial values
\begin{eqnarray}
\nonumber
&\rho_{0,\Lambda}=n_\Lambda=\theta(\sigma_0)\sigma_0/\lambda_\Lambda,\quad m_\Lambda^2=-\theta(-\sigma_0)\sigma_0,&\\
&\bar{A}_\Lambda=1,\quad \alpha_\Lambda=-1, \quad \beta_\Lambda=0.&
\end{eqnarray}
We remain with the free microscopic couplings $\lambda_\Lambda$, $S_\Lambda=\bar{S}_\Lambda$, $V_\Lambda=\bar{V}_\Lambda$. The coupling $\lambda_\Lambda$ will be replaced by the scattering length $a$ in the next section. We further choose units for $\tau$ where $S_\Lambda=1.$ Then our second free coupling is
\begin{equation}
\tilde{v}=\frac{V_\Lambda\Lambda^2}{S_\Lambda^2}=V_\Lambda \Lambda^2.
\end{equation}
In consequence, besides the thermodynamic variables $T$ and $\sigma_0$ our model is characterized by two free parameters, $a$ and $\tilde{v}$. Often, we will concentrate on "standard" non-relativistic bosons with a linear $\tau$ derivative, such that $\tilde{v}=0$. The scattering length $a$ remains then the only free parameter. In the vacuum, where $T=n=0$, this sets the relevant unit of length.

Finally, we have for the infrared cutoff
\begin{equation}
\Delta S_k = \int_x \bar{A}\bar{\phi}^*\,r_k(-\Delta)\bar{\phi} = \int_x 
\phi^*\,r_k(-\Delta)\phi.
\end{equation}
We choose the optimized cutoff function \cite{Litim:2001up}
\begin{equation}
r_k(p^2)=(k^2-p^2-m^2)\theta(k^2-p^2-m^2),
\label{eqCutoff}
\end{equation}
where we recall, that $m^2=0$ in the regime with spontaneous symmetry breaking.
It is convenient to work with real fields $\phi_{1,2}(x)$, $\phi(x)=\frac{1}{\sqrt{2}}(\phi_1(x)+i\phi_2(x))$, with Fourier components \begin{equation}
\phi_j(\tau,\vec{x})=\int_q e^{iqx} \phi_j(q)=\int_{q_0}\int_{\vec{q}} e^{i(q_0\tau+\vec{q}\vec{x})}\phi_j(q_0,\vec{q}).
\end{equation}
The inverse propagator for the fields $\bar{\phi}$ becomes a $2\times2$ matrix in the space of $\bar{\phi}_1$ and $\bar{\phi}_2$, given by the second functional derivative of $\Gamma_k$. For a real constant background field $\bar{\phi}_1(x)=\sqrt{2\bar{\rho}}$, $\bar{\phi}_2(x)=0$ the latter becomes diagonal in momentum space
\begin{equation}
\Gamma_k^{(2)}(q,q^\prime)=G_k^{-1}(q)\delta(q-q^\prime).
\end{equation}
For our truncation one has at $\sigma=\sigma_0$
\begin{equation}
G^{-1}=\bar{A}\begin{pmatrix} \vec{q}^2+V q_0^2+U^\prime+2\rho U^{\prime\prime} & , & -S q_0 \\ S q_0 & , & \vec{q}^2+V q_0^2+U^\prime \end{pmatrix}.
\label{eqprop}
\end{equation}
Here, primes denote derivatives with respect to $\rho$ (not $\bar{\rho}$). In the phase with spontaneous symmetry breaking, the infrared cutoff in the flow equation \eqref{eqFlowequation} adds to the diagonal term in \eqref{eqprop} a piece $\bar{A}(k^2-\vec{q}^2)\theta(k^2-\vec{q}^2)$. This effectively replaces $\vec{q}^2\rightarrow k^2$ in eq. \eqref{eqFlowequation} whenever $\vec{q}^2<k^2$, thus providing for an efficient infrared regularization.

\subsection{Non-perturbative flow equations}
\begin{figure}
\includegraphics{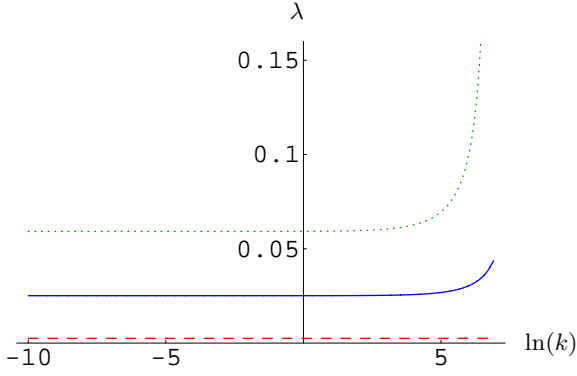}
\caption{(Color online) Flow of the interaction strength $\lambda$ with the scale parameter $k$ for different start values, corresponding to $\lambda_\Lambda=10^3$ (dotted), $\lambda_\Lambda=0.044$ (solid) and $\lambda_\Lambda=0.0026$ (dashed). The first case is plotted for zero density ($n=0$) only, while the last two cases are plotted also for unit density ($n=1$). The curves $n=0$ and $n=1$ are identical within the plot resolution. The solid and the dashed curve correspond to $a=10^{-3}$ and $a=10^{-4}$, respectively.}
\label{figflowoflambda}
\end{figure}

\begin{figure}
\includegraphics{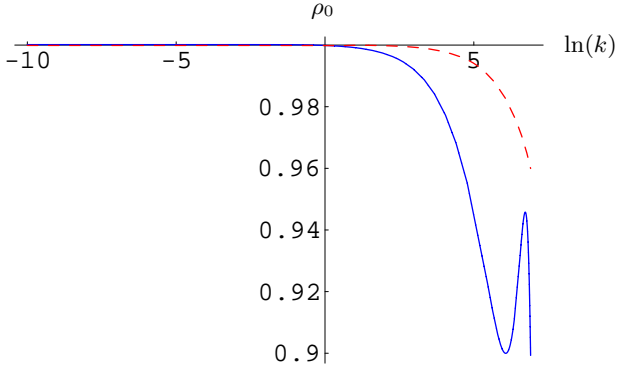}
\caption{(Color online) Flow of the minimum of the effective potential for $n=1$. The parameters for the solid and the dashed curves are the same as in figure \ref{figflowoflambda}.}
\label{figflowofrho}
\end{figure}

We project the flow equation of the effective average action onto equations for the coupling constants by using appropriate background fields and taking functional derivatives. The flow equation for the effective potential obtains by using a space- and time-independent background field in eq. \eqref{eqFlowequation}, with $t=\text{ln}(k/\Lambda)$
\begin{equation}
\partial_t U\big{|}_{\bar{\rho}}=k\,\partial_k U\big{|}_{\bar{\rho}}=\zeta=\frac{1}{2}\int_q \text{tr} \,G_k \,\partial_t(\bar{A}\,r_k).
\label{eqFlowpotentialMatrix}
\end{equation}
The propagator $G_k$ is here determined from
\begin{equation}
G_k^{-1}=G^{-1}+\bar{A}\,\begin{pmatrix} r_k & 0 \\ 0 & r_k \end{pmatrix}=\bar{A}\,\begin{pmatrix} \tilde{P}_{11}, & \tilde{P}_{12} \\ \tilde{P}_{21}, & \tilde{P}_{22} \end{pmatrix},
\end{equation}
with
\begin{eqnarray}
\nonumber
\tilde{P}_{11}&=&k^2+V q_0^2+U^\prime+2\rho U^{\prime\prime}+2V(\sigma-\sigma_0)\vec{q}^2,\\
\nonumber
\tilde{P}_{21}&=&-\tilde{P}_{12}=S q_0+2V(\sigma-\sigma_0)q_0,\\
\tilde{P}_{22}&=&k^2+V q_0^2+U^\prime+2V(\sigma-\sigma_0)\vec{q}^2.
\end{eqnarray}
Again primes denote a differentiation with respect to $\rho$. We switch to renormalized fields by making a change of variables in the differential equation \eqref{eqFlowpotentialMatrix}
\begin{equation}
\partial_t U\big{|}_\rho=\zeta+\eta\rho U^\prime.
\label{eqFlowpotentialrenorm}
\end{equation}
We can now derive the flow equations for the couplings $\rho_0(k)$ and $\lambda(k)$ by appropriate differentiation of \eqref{eqFlowpotentialrenorm} with respect to $\rho$. The flow equation for $U$ is given more explicitly in appendix \ref{secFlowEffectivePotential}. Differentiation with respect to $\sigma$ yields the flow of $n_k$, $\alpha$, $\beta$. We use in detail
\begin{eqnarray}
\nonumber
\frac{d}{dt}\lambda &=& \frac{d}{dt}(\partial_\rho^2U)(\rho_0,\sigma_0)\\
\nonumber
&=& (\partial_\rho^2\partial_tU)(\rho_0,\sigma_0)+(\partial_\rho^3U)(\rho_0,\sigma_0) \,\partial_t\rho_0,\\
&=& \partial_\rho^2\zeta \big{|}_{\rho_0,\sigma_0}+2\eta \lambda,
\end{eqnarray}
where we recall, that $\partial_\rho^3 U=0$ in our truncation. To determine the flow equation of $\rho_0$, we use the condition that $U^\prime(\rho_0)=0$ for all $k$, and therefore
\begin{eqnarray}
\nonumber
&\frac{d}{dt}(\partial_\rho U)(\rho_0,\sigma_0)=0,&\\
\nonumber
&(\partial_\rho\partial_tU)(\rho_0,\sigma_0)+(\partial_\rho^2 U)(\rho_0,\sigma_0)\,\partial_t\rho_0=0,&\\
&\partial_t\rho_0=-\frac{1}{\lambda}(\partial_\rho\partial_t U)(\rho_0,\sigma_0)=-\eta\rho_0-\frac{1}{\lambda}\partial_\rho \zeta\big{|}_{\rho_0,\sigma_0}.&
\end{eqnarray}
We show the flow of $\lambda$ and $\rho_0$ in figs. \ref{figflowoflambda}, \ref{figflowofrho} for $n=1$, $T=0$ and different values of $\lambda_\Lambda$ (with $\tilde{v}=0$). The change in $\rho_0$ is rather modest. This will be different for nonzero temperature.

The flow of $n_k$ is given by
\begin{eqnarray}
\nonumber
\frac{d}{dt}n_k&=&\frac{d}{dt}(-\partial_\sigma U)(\rho_0,\sigma_0)\\
\nonumber
&=& -(\partial_\sigma\partial_t U)(\rho_0,\sigma_0)-(\partial_\rho\partial_\sigma U)(\rho_0,\sigma_0)\,\partial_t\rho_0\\
&=& -\partial_\sigma\zeta\big{|}_{\rho_0,\sigma_0}-\alpha \,\partial_t\rho_0,
\label{eqFlowprescriptionnk}
\end{eqnarray}
and similar for the flow of $\alpha$ and $\beta$,
\begin{eqnarray}
\nonumber
\frac{d}{dt}\alpha&=&\frac{d}{dt}(\partial_\rho\partial_\sigma U)(\rho_0,\sigma_0)\\
\nonumber
&=& (\partial_\rho\partial_\sigma\partial_t U)(\rho_0,\sigma_0)+(\partial_\rho^2\partial_\sigma U)(\rho_0,\sigma_0)\,\partial_t\rho_0\\
\nonumber
&=& \partial_\rho\partial_\sigma\zeta\big{|}_{\rho_0,\sigma_0}+\eta \alpha+\beta (\eta \rho_0+\partial_t\rho_0),\\
\nonumber
\frac{d}{dt}\beta &=&\frac{d}{dt}(\partial_\rho^2\partial_\sigma U)(\rho_0,\sigma_0)\\
\nonumber
&=& (\partial_\rho^2\partial_\sigma\partial_t U)(\rho_0,\sigma_0)+(\partial_\rho^3\partial_\sigma U)(\rho_0,\sigma_0)\partial_t \rho_0,\\
&=& \partial_\rho^2\partial_\sigma \zeta\big{|}_{\rho_0,\sigma_0}+2\eta \beta,
\label{eqflowalphabeta}
\end{eqnarray}
where the last equation holds since $\partial_\rho^3\partial_\sigma U=0$ in our truncation.

\begin{figure}
\includegraphics{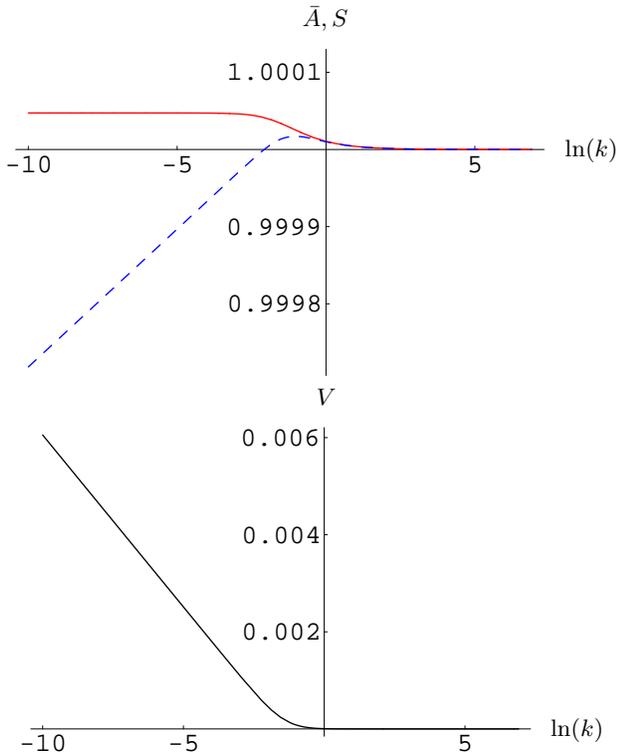}
\caption{(Color online) Flow of the kinetic coefficients $\bar{A}$ (solid), $S$ (dashed) and $V$ for a scattering length $a=10^{-3}$, temperature $T=0$ and density $n=1$.}
\label{figFlowKinetic}
\end{figure}

For a derivation of $\eta=-(\partial_t  \bar{A})/\bar{A}$ and the flow equations for $S$ and $V$, we have to evaluate the flow equation \eqref{eqFlowequation} for a background field depending on $q_0$ and $\vec{q}$. We use an analytic continuation $q_0=i\omega$ and obtain the flow equation for $S$ from
\begin{equation}
\partial_t(S\bar{A})=-i\Omega^{-1}\frac{\partial}{\partial \omega}\frac{\delta}{\delta \bar{\phi}_2(-\omega,0)}\frac{\delta}{\delta \bar{\phi}_1(\omega,0)}\partial_t \Gamma_k\bigg{|}_{\omega=0},
\label{eqflowprescriptionS}
\end{equation}
with four-volume $\Omega=\frac{1}{T}\int_{\vec{x}}$. The projection prescription for $V$ is
\begin{equation}
\partial_t (V\bar{A})=-\Omega^{-1}\frac{\partial}{\partial \omega^2}\frac{\delta}{\delta \bar{\phi}_2(-\omega,0)}\frac{\delta}{\delta \bar{\phi}_2(\omega,0)}\partial_t \Gamma_k\bigg{|}_{\omega=0},
\label{eqflowprescriptionV}
\end{equation}
and similar for $\bar{A}$
\begin{equation}
\partial_t\bar{A}=\Omega^{-1}\frac{\partial}{\partial \vec{p}^2}\frac{\delta}{\delta \bar{\phi}_2(0,-\vec{p})}\frac{\delta}{\delta \bar{\phi}_2(0,\vec{p})}\partial_t \Gamma_k\bigg{|}_{\vec{p}^2=0}.
\label{eqflowprescriptionA}
\end{equation}
After the functional differentiation, we evaluate the expressions \eqref{eqflowprescriptionS}, \eqref{eqflowprescriptionV}, \eqref{eqflowprescriptionA} at homogeneous background fields. These calculations are a little intricate, but standard and straightforward in principle. More explicit flow equations are given in app. \ref{sectFlowofkineticcoefficients}. Eventually, it is always possible to perform the Matsubara sums and also the spatial momentum integration analytically. In figure \ref{figFlowKinetic} we show the flow of $\bar{A}$, $S$ and $V$ at zero temperature and for density $n=1$. 
The kinetic coefficient $\bar{A}$ starts on the large scale with $\bar{A}=1$, increases a little around $k=n^{1/3}$ and saturates to a constant. In contrast, the coefficient $S$ starts to decrease after a short period of increase with $\bar{A}$. For very tiny scales $k$, $S$ would finally go to zero. The frequency dependence of the propagator is then governed by the quadratic frequency coefficient $V$. In three spatial dimensions, however, this decrease of $S$ is so slow that it is not relevant on the length scales of experiments. This is one of the reasons why Bogoliubov theory, which neglects the appearance of $V$, describes experiments with ultracold bosonic quantum gases in three dimensions with so much success. The coefficient $V$ is always generated in the phase with spontaneous symmetry breaking \cite{Wetterich:2007ba}. Its $k$-dependence is also shown in figure \ref{figFlowKinetic}.

\section{Upper bound for the scattering length}
\label{sectUpperboundforthescatteringlength}
The vacuum is defined to have zero temperature $T=0$ and vanishing density $n=0$, which also implies $\rho_0=0$. The interaction strength $\lambda$ at the scale $k=0$ determines the four point vertex at zero momentum. It is directly related to the scattering length $a$ for the scattering of two particles in vacuum, which is experimentally observable. We therefore want to replace the microscopic coupling $\lambda_\Lambda$ by the renormalized coupling $a$. In our units ($2M=1$), one has the relation 
\begin{equation}
a=\frac{1}{8\pi}\lambda(k=0, T=0, n=0).
\end{equation}
The vacuum properties can be computed by taking for $T=0$ the limit $n\rightarrow0$. We may also perform an equivalent and technically simple computation in the symmetric phase by choosing $m^2(k=\Lambda)$ such, that $m^2(k\rightarrow0)=0$. This guarantees that the boson field $\phi$ is a gap-less propagating degree of freedom. 

We first investigate the model with a linear $\tau$-derivative, $S_\Lambda=1$, $V_\Lambda=0$. Projecting the flow equation \eqref{eqFlowequation}, we find the following equations:
\begin{eqnarray}
\nonumber \partial_t m^2 & = & 0\\
\partial_t \lambda & = & \left(\frac{\lambda^2}{6}\right)\frac{{\left( k^2 - m^2 \right) }^{3/2}}{k^2\,
    {\pi }^2\,S}\,\Theta(k^2-m^2).
\label{eqflowvacuummlambda}
\end{eqnarray}
The propagator is not renormalized, $\partial_t S=\partial_tV=\partial_t \bar{A}=0$, $\eta=0$, $\partial_t\alpha=0$, and one finds $\partial_tn_k=0$. The coupling $\beta$ is running according to
\begin{eqnarray}
\nonumber \partial_t \beta & = & \left(\frac{1}{3}\alpha\lambda^2-\frac{1}{3}k^2\beta\lambda\right)\\
&&\frac{\left(k^2 - m^2\right)^{3/2}}{k^4\,\pi^2 \,S}\Theta(k^2 - m^2).
\label{eqflowvacuumbeta}
\end{eqnarray}
Since $\beta$ appears only in its own flow equation, it is of no further relevance in the vacuum. Also, no coupling $V$ is generated by the flow and we have therefore set $V=0$ on the r.h.s. of eqs. \eqref{eqflowvacuummlambda} and \eqref{eqflowvacuumbeta}. 

\begin{figure}
\includegraphics{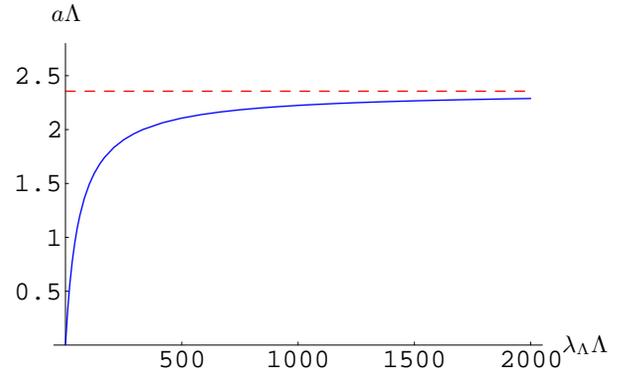}
\caption{(Color online) Scatt\-er\-ing length $a$ in dependence on the microscopic interaction strength $\lambda_\Lambda$ (solid). The asymptotic maximum $a_{\text{max}}=\frac{3 \pi}{4\Lambda}$ is also shown (dashed).}
\label{figscatteringbound}
\end{figure}

Inserting in eq. \eqref{eqflowvacuummlambda} the vacuum values $m^2=0$ and $S=1$, we find
\begin{equation}
\partial_t\lambda=\frac{k}{6\pi^2}\lambda^2.
\end{equation}
The solution 
\begin{equation}
\lambda(k)=\frac{1}{\frac{1}{\lambda_\Lambda}+\frac{1}{6\pi^2}(\Lambda-k)}
\end{equation}
tends to a constant for $k\rightarrow0$, $\lambda_0=\lambda(k=0)$. The dimensionless variable $\tilde{\lambda}=\frac{\lambda k}{S}$ goes to zero, when $k$ goes to zero. This shows the infrared freedom of the theory. For fixed ultraviolet cutoff, the scattering length
\begin{equation}
a=\frac{\lambda_0}{8\pi}=\frac{1}{\frac{8\pi}{\lambda_\Lambda}+\frac{4}{3\pi}\Lambda},
\end{equation}
as a function of the initial value $\lambda_\Lambda$, has an asymptotic maximum
\begin{equation}
a_{\text{max}}=\frac{3\pi}{4\Lambda}.
\label{eqscatteringbound}
\end{equation}
The relation between $a$ and $\lambda_\Lambda$ is shown in fig. \ref{figscatteringbound}.

As a consequence of eq. \eqref{eqscatteringbound}, the nonrelativistic bosons in $d=3$ are a "trivial theory" in the sense that the bosons become noninteracting in the limit $\Lambda\rightarrow\infty$, where $a\rightarrow0$. The upper bound \eqref{eqscatteringbound} has important practical consequences. It tells us, that whenever the "macrophysical length scales" are substantially larger than the microscopic length $\Lambda^{-1}$, we deal with a weakly interacting theory. As an example, consider a boson gas with a typical inter-particle distance substantially larger than $\Lambda^{-1}$. (For atom gases $\Lambda^{-1}$may be associated with the range of the Van der Waals force.) We may set the units in terms of the particle density $n$, $n=1$. In these units $\Lambda$ is large, say $\Lambda=10^3$. This implies a very weak interaction, $a\lesssim 2.5\cdot10^{-3}$. In other words, the scattering length cannot be much larger than the microscopic length $\Lambda^{-1}$. For such systems, perturbation theory will be valid in many circumstances. We will find that the Bogoliubov theory indeed gives a reliable account of many properties. Even for an arbitrary large microphysical coupling $(\lambda_\Lambda\rightarrow\infty)$, the renormalized physical scattering length $a$ remains finite.

Let us mention, however, that the weak interaction strength does not guarantee the validity of perturbation theory in all circumstances. For example, near the critical temperature of the phase transition between the superfluid and the normal state, the running of $\lambda(k)$ will be different from the vacuum. As a consequence, the coupling will vanish proportional to the inverse correlation length $\xi^{-1}$ as $T$ approaches $T_c$, $\lambda \sim T^{-2}\xi^{-1}$. Indeed, the phase transition will be characterized by the non-perturbative critical exponents of the Wilson-Fisher fixed point. Also for lower dimensional systems, the upper bound \eqref{eqscatteringbound} for $\lambda_0$ is no longer valid - for example the running of $\lambda$ is logarithmic for $d=2$. For our models with $V_\Lambda\neq0$, the upper bound becomes dependent on $V_\Lambda$. It increases for $V_\Lambda>0$. In the limit $S_\Lambda\rightarrow0$, it is replaced by the well known "triviality bound" of the four dimensional relativistic model, which depends only logarithmically on $\Lambda$. Finally, for superfluid liquids, as $^4\text{He}$, one has $n\sim\Lambda^3$, such that for $a\sim \Lambda^{-1}$ one finds a large concentration $c$.

The situation for dilute bosons seems to contrast with ultracold fermion gases in the unitary limit of a Feshbach resonance, where $a$ diverges. One may also think about a Feshbach resonance for bosonic atoms, where one would expect a large scattering length for a tuning close to resonance. In this case, however, the effective action does not remain local. It is best described by the exchange of molecules. The scale of nonlocality is then given by the gap for the molecules, $m_M$. Only for momenta $\vec{q}^2<m_M^2$ the effective action becomes approximately local, such that $\Lambda=m_M$ for our approximation. Close to resonance, the effective cutoff is low and again in the vicinity of $a^{-1}$.

\section{Quantum phase diagram}
\label{sectQuantumphasediagram}
\subsection{Different methods to determine the density}
The density sets a crucial scale for our problem. Its precise determination is mandatory for quantitative precision. We will discuss two different methods for its determination and show that the results agree within our precision. For $T=0$, we also find agreement with the Ward identity $n=\rho_0$.

The first method is to derive flow equations for the density. This has the advantage that the occupation numbers for a given momentum $\vec{p}$ are mainly sensitive to running couplings with $k^2=\vec{p}^2$. In the grand canonical formalism, the density is defined by
\begin{equation}
n=-\frac{\partial}{\partial \sigma}\frac{1}{\Omega}\Gamma[\phi]{\Big |}_{\phi=\phi_0,\sigma=\sigma_0}
\end{equation}
We can formally define a $k$-dependent density $n_k$ by
\begin{equation}
n_k=-\frac{\partial}{\partial \sigma}\frac{1}{\Omega}\Gamma_k[\phi]{\Big |}_{\phi=\phi_0,\sigma=\sigma_0}=-(\partial_\sigma U)(\rho_0,\sigma_0).
\end{equation}
The flow equation for $n_k$ is given in eq. \eqref{eqFlowprescriptionnk} and the physical density obtains for $k=0$. The term $\partial_\sigma \zeta \big{|}_{\rho_0,\sigma_0}$ that enters eq. \eqref{eqFlowprescriptionnk} is the derivative of the flow equation \eqref{eqFlowpotentialMatrix} for $U$ with respect to $\sigma$. To compute it, we need the $\sigma$-dependence of the propagator $G_k$ in the vicinity of $\sigma_0$. Within a systematic derivative expansion, we use the expansion of $U(\rho,\sigma)$ and the kinetic coefficients $Z_1$ and $Z_2$ to linear order in $(\sigma-\sigma_0)$, as described in appendix \ref{sectTruncation}. Here, $Z_1(\rho,\sigma)$ and $Z_2(\rho,\sigma)$ are the coefficient functions of the terms linear in the $\tau$-derivative and linear in $\Delta$, respectively. 
No reasonable qualitative behavior is found, if the linear dependence of $Z_1$ and $Z_2$ on $(\sigma-\sigma_0)$ is neglected. Also, the scale dependence of $\alpha$ and $\beta$ are quite important. The flow equations for $\alpha$ and $\beta$ can be obtained directly by differentiating the flow equation of the effective potential with respect to $\sigma$ and $\rho$, cf. eq. \eqref{eqflowalphabeta}. The situation is more complicated for the kinetic coefficients $Z_1^{(\sigma)}=\partial_\sigma Z_1(\rho_0,\sigma_0)$ and $Z_2^{(\sigma)}=\partial_\sigma Z_2(\rho_0,\sigma_0)$. Their flow equations have to be determined by taking the $\sigma$-derivative of the flow equation for $Z_1(\rho, \sigma)$ and $Z_2(\rho,\sigma)$. As discussed in app. \ref{sectTruncation}, we use in this paper the approximation $Z_1^{(\sigma)}=Z_2^{(\sigma)}=2V=2V_1(\rho_0,\sigma_0)$.

As a check of both our method and our numerics, we also use another way to determine the particle density. This second method is more robust with respect to shortcomings of the truncation, but less adequate for high precision calculations as needed e.g. to determine the condensate depletion. The second method determines the pressure $p=-U(\rho_0,\sigma_0)$ as a function of the chemical potential $\sigma_0$. Here, the effective potential is normalized by $U(\rho_0=0,\sigma_0)=0$ at $T=0$, $n=0$. The flow of the pressure can be read of directly from the flow equation of the effective potential and is independent of the couplings $\alpha$ and $\beta$. We calculate the pressure as a function of $\sigma$ and determine the density $n=\frac{\partial}{\partial \sigma}p$ by taking the $\sigma$-derivative numerically. It turns out that $p$ is in very good approximation given by $p=c\,\sigma^2$, where the constant $c$ can be determined from a numerical fit. The density is thus linear in $\sigma$.  

\begin{figure}
\includegraphics{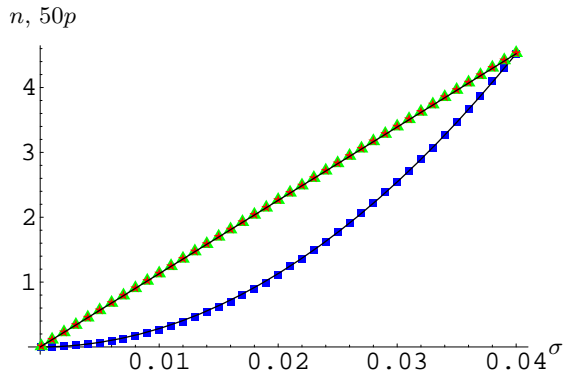}
\caption{(Color online) Pressure and density as a function of the chemical potential at $T=0$. We use three different methods: $n=-\partial_\sigma U_{\text{min}}$ from the flow equation (triangles), $n=\rho_0$ as implied by Galilean symmetry (stars) and $n=\partial_\sigma p$, where the pressure $p=-U$ (boxes) was obtained from the flow equation and phenomenologically fitted by $p=56.5 \sigma^2$ (solid lines). Units are arbitrary and we use $a=3.4\cdot 10^{-4}$, $\Lambda=10^3$.}
\label{densitycompared}
\end{figure}

At zero temperature and for $\tilde{v}=0$, we can additionally use the Ward identities connected to Galilean symmetry, which yield $n=\rho_0$. We compare our methods in figure \ref{densitycompared} and find that they give numerically the same result. We stress again the importance of a reliable method to determine the density, since we often rescale variables by powers of the density to obtain dimensionless variables.

\subsection{Condensate and depletion density for $T=0$}

We want to split the density into a condensate part $n_c$ and a density for uncondensed particles or "depletion density" $n_d=n-n_c$. For our model the condensate density is given by the "bare" order parameter
\begin{equation}
n_c=\bar{\rho}_0=\bar{\rho}_0(k=0).
\end{equation}
In order to show this, we introduce occupation numbers $n(\vec{p})$ for the modes with momentum $\vec{p}$ with normalization
\begin{equation}
\int_{\vec{p}}n(\vec{p})=n.
\end{equation}
One formally introduces a $\vec{p}$ dependent chemical potential $\sigma(\vec{p})$ in the grand canonical partition function
\begin{equation}
e^{-\Gamma_{\text{min}}[\sigma]}=\text{Tr}e^{-\beta(H-\Omega_3\int_{\vec{p}}\sigma(\vec{p})n(\vec{p}))},
\end{equation}
with three dimensional volume $\Omega_3=\int_{\vec{x}}$. Then one can define the occupation numbers by
\begin{equation}
n(\vec{p})=-\frac{\delta}{\delta \sigma(\vec{p})}\frac{1}{\beta \Omega_3}\Gamma[\phi,\sigma(\vec{p})]{\Big |}_{\phi=\phi_0,\sigma(\vec{p})=\sigma_0}.
\end{equation}
This construction allows us to use $k$-dependent occupation numbers by the definition
\begin{equation}
n_k(\vec{p})=-\frac{\delta}{\delta \sigma(\vec{p})}\frac{1}{\beta \Omega_3}\Gamma_k[\phi,\sigma]{\Big |}_{\phi=\phi_0,\sigma(\vec{p})=\sigma_0}.
\end{equation}
One can derive a flow equation for this occupation number $n_k(\vec{p})$ \cite{Wetterich:OccupationNumbers}:
\begin{eqnarray}
\nonumber
\partial_k n_k(\vec{p}) &=& -\frac{1}{2}\frac{\delta}{\delta \sigma(\vec{p})}\frac{1}{\beta \Omega_3}\text{Tr}\left\{(\Gamma^{(2)}+R_k)^{-1}\partial_k R_k\right\}\big{|}_{\phi_0,\sigma_0}\\
& & +\frac{\partial}{\partial \rho}\frac{\delta}{\delta \sigma(\vec{p})}\frac{1}{\beta \Omega_3}\Gamma[\phi,\sigma]{\Big |}_{\phi_0,\sigma_0}(\partial_k\rho_0).
\label{flowofnp}
\end{eqnarray}

We split the density occupation number into a $\delta$-distribution like part and a depletion part, which is regular in the limit $\vec{p}\rightarrow0$
\begin{equation}
n_k(\vec{p})=n_{c,k}\,\delta(\vec{p})+n_{d,k}(\vec{p}).
\end{equation}
One can see from the flow equation for $n_k(\vec{p})$ that the only contribution to $\partial_k n_{c,k}$ comes from the second term in equation \eqref{flowofnp}. Within a more detailed analysis \cite{Wetterich:OccupationNumbers} one finds
\begin{equation}
\partial_k n_{c,k}=\partial_k\bar{\rho}_{0,k}.
\end{equation}
We therefore identify the condensate density with the bare order parameter
\begin{equation}
n_c=\bar{\rho}_0=\frac{\rho_0}{\bar{A}}=\bar{\phi}_0^2.
\end{equation}
Correspondingly, we define the $k$-dependent quantities
\begin{equation}
n_{c,k}=\bar{\rho}_{0,k},\quad n_k=n_{c,k}+n_{d,k}
\end{equation}
and compute $n_d=n_d(k=0)$ by a solution of its flow equation. 

Even at zero temperature, the repulsive interaction connected with a positive scattering length $a$ causes a portion of the particle density to be outside the condensate. From dimensional reasons, it is clear, that $n_d/n=(n-n_c)/n$ should be a function of $an^{1/3}$. The prediction of Bogoliubov theory or, equivalently, mean field theory, is $n_d/n=\frac{8}{3\sqrt{\pi}}(an^{1/3})^{3/2}$. We may determine the condensate depletion from the solution to the flow equation for the particle density, $n=n_{k=0}$, and $n_c=\bar{\rho}_0=\bar{\rho}_0(k=0)$. 

\begin{figure}
\includegraphics{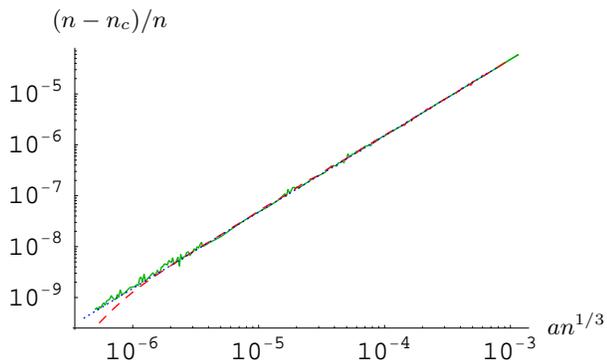}
\caption{(Color online) Condensate depletion $(n-n_c)/n$ as a function of the dimensionless scattering length $a n^{1/3}$. For the solid curve, we vary $a$ with fixed $n=1$, for the dashed curve we vary the density at fixed $a=10^{-4}$. The dotted line is the Bogoliubov-Result $(n-n_c)/n=\frac{8}{3\sqrt{\pi}}(a n^{1/3})^{3/2}$ for reference. We find perfect agreement of the three determinations. The fluctuations in the solid curve for small $a n^{1/3}$ are due to numerical uncertainties. Their size demonstrates our numerical precision.}
\label{figDepletion}
\end{figure}

From Galilean invariance for $T=0$ and $\tilde{v}=0$, it follows that
\begin{equation}
\frac{n_d}{n}=\frac{\rho_0-\bar{\rho}_0}{\rho_0}=1-\frac{1}{\bar{A}},
\end{equation}
with $\bar{A}=\bar{A}(k=0)$. This gives an independent determination of $n_c$. 
In figure \ref{figDepletion} we plot the depletion density obtained from the flow of $n$ and $\bar{\rho}_0$ over several orders of magnitude. Apart from some numerical fluctuations for small $an^{1/3}$, we find that our result is in full agreement with the Bogoliubov prediction.

\subsection{Quantum phase transition}
For $T=0$ a quantum phase transition separates the phases with $\rho_0=0$ and $\rho_0>0$.
In this section, we investigate the phase diagram at zero temperature in the cube spanned by the dimensionless parameters $\tilde{\sigma}=\frac{\sigma}{\Lambda^2}$, $\tilde{a}=a\Lambda$ and $\tilde{v}=\frac{V_\Lambda}{S_\Lambda^2}\Lambda^2$. This goes beyond the usual phase transition for nonrelativistic bosons, since we also include a microscopic second $\tau$-derivative $\sim\tilde{v}$, and therefore models with a generalized microscopic dispersion relation.
For non-vanishing $\tilde{v}$ (i.e. for a nonzero initial value of $V_1$ with $V_2=V_3=0$ in app. \ref{sectTruncation}), the Galilean invariance at zero temperature is broken explicitly. For large $\tilde{v}$, we expect a crossover to the "relativistic" $O(2)$ model. If we send the initial value of the coefficient of the linear $\tau$-derivative $S_\Lambda$ to zero, we obtain the limiting case $\tilde{v}\rightarrow\infty$. The symmetries of the model are now the same as those of the relativistic $O(2)$ model in four dimensions. The space-time-rotations or Lorentz symmetry replace Galilean symmetry. 

\begin{figure}
\includegraphics{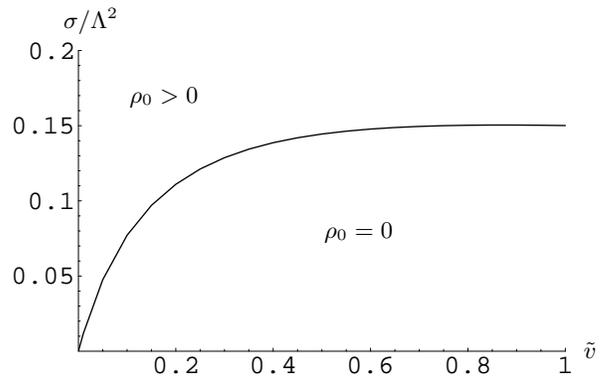}
\caption{Quantum phase diagram in the $\tilde{\sigma}$-$\tilde{v}$ plane for $\tilde{a}=1$.}
\label{figQPTsigmava1}
\end{figure}

It is interesting to study the crossover between the two cases. Since our cutoff explicitly breaks Lorentz symmetry, we investigate in this paper only the regime $\tilde{v}\lesssim1$. Detailed investigations of the flow equations for $\tilde{v}\rightarrow\infty$ can be found in the literature \cite{Papenbrock:1994kf, Berges:2000ew, CE}. The phase diagram in the $\tilde{\sigma}-\tilde{v}$ plane with $\tilde{a}=1$ is shown in figure \ref{figQPTsigmava1}. The critical chemical potential first increases linearly with $\tilde{v}$ and then saturates to a constant. The slope in the linear regime as well as the saturation value depend linearly on $\tilde{a}$ for $\tilde{a}<1$. 

At $T=0$, the critical exponents are everywhere the mean field ones ($\eta=0$, $\nu=1/2$). This is expected: It is the case for $\tilde{v}=0$ \cite{Wetterich:2007ba, CIPT}, and for $\tilde{v}=\infty$ the theory is equivalent to a relativistic $O(2)$ model in $d=3+1$ dimensions. This is just the upper critical dimension of the Wilson-Fisher fixed point \cite{WilsonFisher}. 

\begin{figure}
\includegraphics{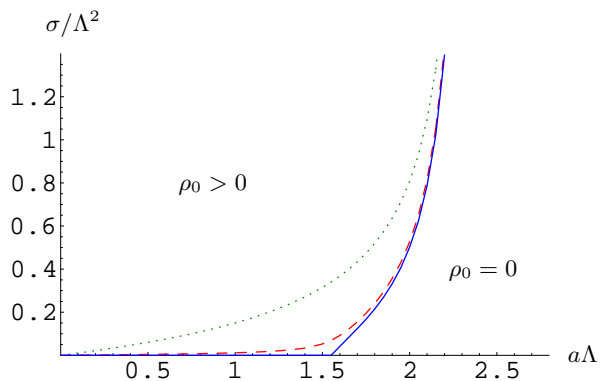}
\caption{(Color online) Quantum phase diagram in the $\tilde{\sigma}$-$\tilde{a}$ plane for $\tilde{v}=1$ (dotted), $\tilde{v}=0.01$ (dashed) and $\tilde{v}=0$ (solid).}
\label{figQPTsigmaofa}
\end{figure}

From section 3 we know that for $\tilde{v}=0$ the parameter $\tilde{a}$ is limited to $\tilde{a}<\frac{3\pi}{4}\approx 2.356$. For $\tilde{v}=0$ and a small scattering length $a\rightarrow0$, a second order quantum phase transition divides the phases without spontaneous symmetry breaking  for $\sigma<0$ from the phase with a finite order parameter $\rho_0>0$ for $\sigma>0$. It is an interesting question, whether this quantum phase transition at $\sigma=0, \tilde{v}=0$ also occurs for larger scattering length $a$. We find in our truncation that this is indeed the case for a large range of $a$, but  not for $\tilde{a}>1.55$. Here, the critical chemical potential suddenly increases to large positive values as shown in fig. \ref{figQPTsigmaofa}. For $\tilde{v}>0$ this increase happens even earlier. (For a truncation with $V_1\equiv0$, the phase transition would always occur at $\sigma=0$.) We plot the $\tilde{\sigma}-\tilde{a}$ plane of the phase diagram for different values of $\tilde{v}$ in figure \ref{figQPTsigmaofa}. The form of the critical line can be understood by considering the limits $\tilde{v}\rightarrow0$ as well as $\tilde{a}\rightarrow0$.  

\begin{figure}
\includegraphics{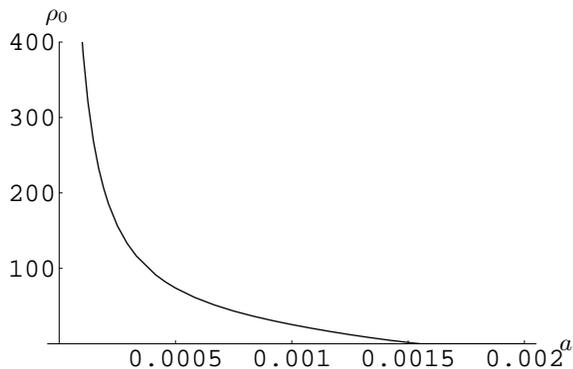}
\caption{Quantum phase transition for fixed chemical potential $\sigma=1$, with $\Lambda=10^3$. The density $\rho_0=n$ as a function of the scattering length $a$ goes to zero at a critical $a_c \Lambda=1.55$, indicating a second order quantum phase transition at that point.}
\label{figQPTrhoofa}
\end{figure}

For a fixed chemical potential, the order parameter $\rho_0$ as a function of $a$ goes to zero at a critical value $a_c$ as shown in fig. \ref{figQPTrhoofa}. This happens in a continuous way and the phase transition is therefore of second order.  For $\sigma\rightarrow0$, we find $a_c=1.55 \Lambda^{-1}$. We emphasize, however, that $a_c$ is of the order of the microscopic distance $\Lambda^{-1}$. Universality may not be realized for such values, and the true phase transition may depend on the microphysics. For example, beyond a critical value for the repulsive interaction, the system may form a solid. Ultracold atom gases correspond to metastable states which may lose their relevance for $a\rightarrow\Lambda^{-1}$. For $\tilde{v}>0$ and $\sigma\ll\Lambda^2$ the phase transition occurs for $a_c \Lambda \ll 1$ such that universal behavior is expected.

\section{Temperature dependence of condensate}
\label{sectTemperaturedependenceofcondensate}
So far, we have only discussed the vacuum and the dense system at zero temperature. A non vanishing temperature $T$ will introduce an additional scale in our problem. For small $T\ll n^{2/3}$ we expect only small corrections. However, as $T$ increases the superfluid order will be destroyed. Near the phase transition for $T \approx T_c$ and for the disordered phase for $T>T_c$, the characteristic behavior of the boson gas will be very different from the $T\rightarrow0$ limit.

\begin{figure}
\includegraphics{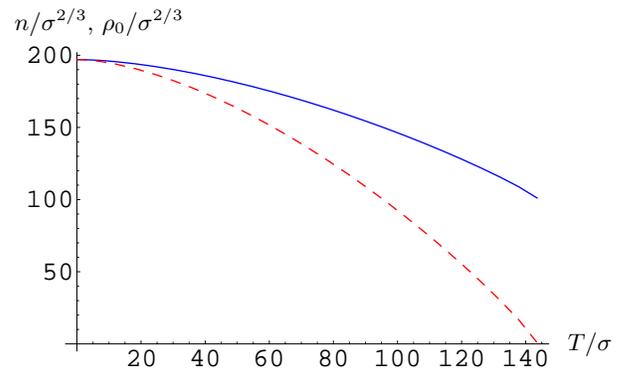}
\caption{(Color online) Density $n/{\sigma^{2/3}}$ (solid) and order parameter $\rho_0/{\sigma^{2/3}}$ (dashed) as a function of the temperature $T/\sigma$. The units are arbitray with $a=2\cdot 10^{-4}$ and $\Lambda=10^3$. The plot covers only the superfluid phase. For higher temperatures, the density is given by the thermal contribution $n=n_T$ only.}
\label{fignoftemperature}
\end{figure}

For $T>0$ the particle density $n$ receives a contribution from a thermal gas of bosonic (quasi-) particles. It is no longer uniquely determined by the superfluid density $\rho_0$. We may write 
\begin{equation}
n=\rho_0+n_T
\label{eqDensityatT}
\end{equation}
and observe, that $n_T=0$ is enforced by Galilean symmetry only for $T=0, V_\Lambda=0$. The heat bath singles out a reference frame, such that for $T>0$ Galilean symmetry no longer holds. In our formalism, the thermal contribution $n_T$ appears due to modifications of the flow equations for $T\neq0$. We start for high $k$ with the same initial values as for $T=0$. As long as $k\gg \pi T$ the flow equations receive only minor modifications. For $k \approx \pi T$ or smaller, however, the discreteness of the Matsubara sum has important effects. We plot in fig. \ref{fignoftemperature} the density as a function of $T$ for fixed $\sigma=1$.

\begin{figure}
\includegraphics{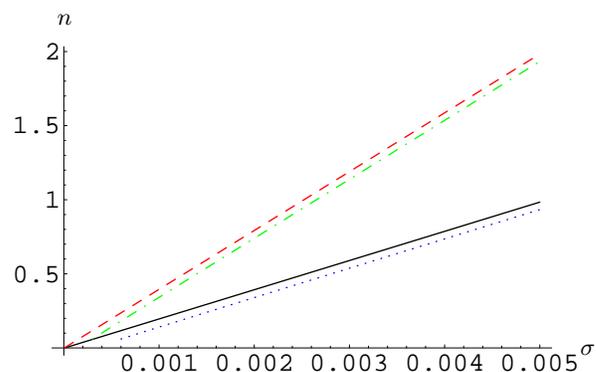}
\caption{(Color online) Density $n$ for different temperatures and scattering length. We plot $n(\sigma)$ in arbitrary units, with $\Lambda=10^3$, and for a scattering length $a=2\cdot10^{-4}$ (solid and dotted), $a=10^{-4}$ (dashed and dashed-dotted). The temperature is $T=0$ (solid and dashed) and $T=1$ (dotted and dashed-dotted).}
\label{fignofsigma}
\end{figure}

In fig. \ref{fignofsigma} we show $n(\sigma)$, similar to fig. \ref{densitycompared}, but now for different $a$ and $T$. For $T=0$ the scattering length sets the only scale besides $n$ and $\sigma$, such that by dimensional arguments $a^2\sigma=f(a^3 n)$. Bogoliubov theory predicts 
\begin{equation}
f(x)=8\pi x(1+\frac{32}{3\sqrt{\pi}}x^{1/2}).
\end{equation}
The first term on the r.h.s. gives the contribution of the ground state, while the second term is added by fluctuation effects. For small scattering length $a$, the ground state contribution dominates. We have then $\sigma\sim a$ for $n=1$ and $\sigma/n$ can be treated as a small quantity. For $T\neq 0$ and small $a$ one expects $\sigma=g(T/n^{2/3})an$. The curves in fig. \ref{fignofsigma} for $T=1$ show that the density, as a function of $\sigma$, is below the curve obtained at $T=0$. This is reasonable, since the statistical fluctuations now drive the order parameter $\rho_0$ to zero. At very small $\sigma$, the flow enters the symmetric phase. The density is always positive, but for simplicity, we show the density as a function of $\sigma$ in figure \ref{fignofsigma} only in those cases, where the flow remains in the phase with spontaneous $U(1)$ symmetry breaking. 

For temperatures above the critical temperature, the order parameter $\rho_0$ vanishes at the macroscopic scale and so does the condensate density $n_c=\bar{\rho_0}=\frac{1}{\bar{A}}\rho_0$. The density is now given by a thermal distribution of particles with nonzero momenta. 
Up to small corrections from the interaction $\sim aT$, it is described by a free Bose gas, 
\begin{equation}
n= \frac{T^{3/2}}{(4\pi)^{3/2}}g_{3/2}(e^{\beta \sigma}),
\end{equation}
with the "Bose function"
\begin{equation}
g_p(z)=\frac{1}{\Gamma(p)}\int_0^\infty dx\,x^{p-1}\frac{1}{z^{-1}e^x-1}.
\end{equation}

\begin{figure}
\includegraphics{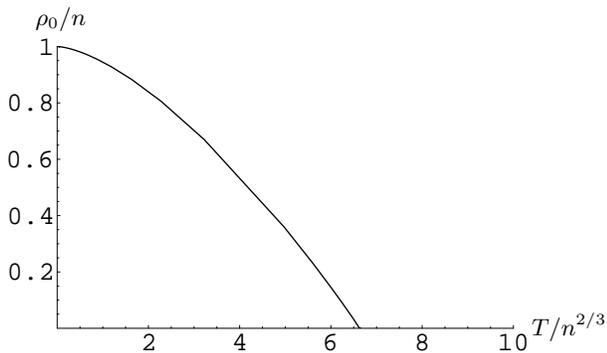}
\caption{Order parameter $\rho_0/n$ as a function of the dimensionless temperature $T/(n^{2/3})$ for scattering length $a=10^{-4}$. Here, we varied $T$ keeping $\sigma$ fixed. Numerically, this is equivalent to varying $\sigma$ with fixed $T$.\label{figrhooftemperature}}
\end{figure}

\begin{figure}
\includegraphics{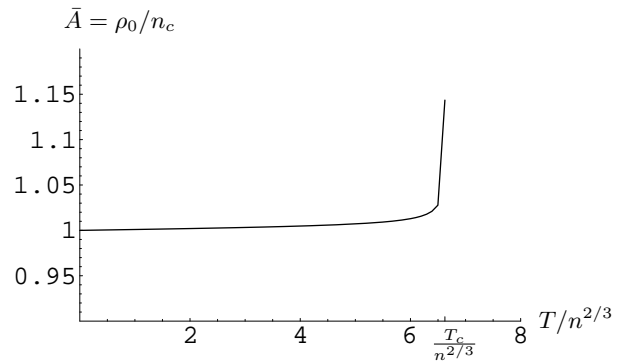}
\caption{Order parameter divided by the condensate density $\bar{A}=\rho_0/n_c$, as a function of the dimensionless temperature $T/(n^{2/3})$, and for scattering length $a=10^{-4}$. Here, we varied $T$ keeping $\sigma$ fixed. Numerically, this is equivalent to varying $\sigma$ with fixed $T$. The plot covers only the phase with spontaneous symmetry breaking. For higher temperatures, the symmetric phase has $\rho_0=n_c=0$. The divergence of $\bar{A}$ for $T\rightarrow T_c$ reflects the anomalous dimension $\eta$ of the Wilson-Fisher fixed point.}
\label{Abaroftemperature}
\end{figure}

In figure \ref{figrhooftemperature} we show the dimensionless order parameter $\rho_0/n$ as a function of the dimensionless temperature $T/n^{2/3}$. 
This plot shows the second order phase transition from the phase with spontaneous $U(1)$ symmetry breaking at small temperatures to the symmetric phase at higher temperatures. The critical temperature $T_c$ is determined as the temperature, where the order parameter just vanishes - it is investigated in the next section. Since we find $(\bar{A}-1)\ll1$, the condensate fraction $n_c/n=\bar{\rho_0}/n=\rho_0/(\bar{A}n)$ as a function of $T/n^{2/3}$ resembles the order parameter $\rho_0/n$. We plot $\bar{A}$ as a function of $T/n^{2/3}$ in fig. \ref{Abaroftemperature}. Except for a narrow region around $T_c$, the deviations from one remain indeed small. Near $T_c$ the gradient coefficient $\bar{A}$ diverges according to the anomalous dimension, $\bar{A}\sim \xi^\eta$, with $\eta$ the anomalous dimension. The correlation length $\xi$ diverges with the critical exponent $\nu$, $\xi\sim |T-T_c|^{-\nu}$, such that
\begin{equation}
\bar{A}\sim |T-T_c|^{-\eta\nu}.
\end{equation}
Here, $\eta$ and $\nu$ are the critical exponents for the Wilson Fisher fixed point of the classical three-dimensional $O(2)$ model, $\eta=0.0380(4)$, $\nu=0.67155(27)$ \cite{Pelissetto:2000ek, Berges:2000ew, CE}.

\begin{figure}
\includegraphics{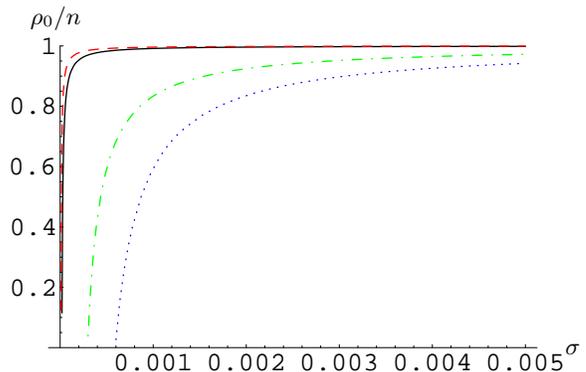}
\caption{(Color online) Order parameter divided by the density, $\rho_0/n$, as a function of the chemical potential. We use arbitrary units with $\Lambda=10^3$. The curves are given for a scattering length $a=2\cdot10^{-4}$ (solid and dotted), $a=10^{-4}$ (dashed and dashed-dotted) and temperature $T=0.1$ (solid and dashed) and $T=1$ (dotted and dashed-dotted). At zero temperature, Galilean invariance implies $\rho_0=n$, which we find within our numerical resolution.}
\label{figrhoofsigma}
\end{figure}

In figure \ref{figrhoofsigma} we plot $\rho_0/n$ as a function of the chemical potential $\sigma$ for different temperatures and scattering lengths. We find, that $\rho_0/n=1$ is indeed approached in the limit $T\rightarrow0$, as required by Galilean invariance. All figures of this section are for $\tilde{v}=0$. The modifications for $\tilde{v}\neq0$ are mainly quantitative, not qualitative. 

\section{Critical temperature}
\label{sectCriticaltemperature}
The critical temperature $T_c$ for the phase transition between the superfluid phase at low temperature and the disordered or symmetric phase at high temperature depends on the scattering length $a$. By dimensional reasoning, the temperature shift $\Delta T_c=T_c(a)-T_c(a=0)$ obeys $\Delta T_c/T_c\sim an^{1/3}$. The proportionality coefficient cannot be computed in perturbation theory \cite{Andersen:2003qj}. It depends on $\tilde{v}$ and we concentrate here on $\tilde{v}=0$. Monte-Carlo simulations in the high temperature limit, where only the lowest Matsubara frequency is included, yield $\Delta T_c/T_c=1.3 \,a n^{1/3}$ \cite{KashurnikovArnold}. Within the same setting, renormalization group studies \cite{Blaizot, Kopietz} yield a similar result, for  composite bosons see \cite{Diehl:2007th}. Near $T_c$, the long wavelength modes with momenta $\vec{p}^2\ll(\pi T)^2$ dominate the "long distance quantities". Then a description in terms of a classical three dimensional system becomes valid. This "dimensional reduction" is achieved by "integrating out" the nonzero Matsubara frequencies. 
%However, the relevant interaction strength in the classical theory, $\lambda_3=8\pi a_3$, still needs to be related to the quantum theory - we need the relation between $a_3$ and $a$. This translation cannot be done within the classical theory. 
However, both $\Delta T_c/T_c$ and $n$ are dominated by modes with momenta $\vec{p}^2\approx(\pi T_c)^2$ such that corrections to the classical result may be expected.

\begin{figure}
\includegraphics{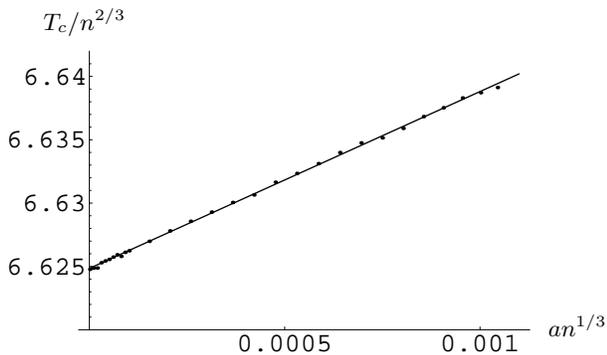}
\caption{Dimensionless critical temperature $T_c/(n^{2/3})$ as a function of the dimensionless scattering length $an^{1/3}$ (points). We also plot the linear fit $\Delta T_c/T_c=2.1\, a n^{1/3}$ (solid line).}
\label{Tcofa}
\end{figure}

We have computed $T_c$ numerically by monitoring the zero of $\rho_0$, as shown in fig. \ref{figrhooftemperature}, $\rho_0(T\rightarrow T_c)\rightarrow0$. Our result is plotted in fig. \ref{Tcofa}. In the limit $a\rightarrow0$ we find for the dimensionless critical temperature $T_c/(n^{2/3})=6.6248$, which is in good agreement with the expected result for the free theory $T_c/(n^{2/3})=\frac{4\pi}{\zeta(3/2)^{2/3}}=6.6250$. For the shift in $T_c$ due to the finite interaction strength, we obtain
\begin{equation}
\frac{\Delta T_c}{T_c}=\kappa \,a n^{1/3}, \quad \kappa=2.1.
\end{equation}
We expect that the result for $\kappa$ depends on the truncation and may change somewhat if additional higher order couplings are included.

\section{Sound velocity}
\label{sectSoundvelocity}
The macroscopic sound velocity $v_S$ is a crucial quantity for the hydrodynamics of the gas or liquid. It is accessible to experiment. As a thermodynamic observable, the adiabatic sound velocity is defined as
\begin{equation}
v_S^2=\frac{1}{M}\frac{\partial p}{\partial n}\bigg{|}_s
\end{equation}
where $M$ is the particle mass (in our units $1/M=2$), $p$ is the pressure, $n$ is the particle density, and $s$ is the entropy per particle. It is related to the isothermal sound velocity $v_T$ by
\begin{equation}
v_S^2=\frac{1}{M}\left(\frac{\partial p}{\partial n}\bigg{|}_T+\frac{\partial p}{\partial T}\bigg{|}_n\frac{\partial T}{\partial n}\bigg{|}_s\right)=v_T^2+2\frac{\partial p}{\partial T}\bigg{|}_n \frac{\partial T}{\partial n}\bigg{|}_s
\label{eqSoundAdiabaticIsothermal}
\end{equation}
where we use our units $2M=1$. One needs the "equation of state" $p(T,n)$ and
\begin{equation}
s(T,n)=\frac{S}{N}=\frac{1}{n}\frac{\partial p}{\partial T}\bigg{|}_\sigma.
\end{equation}
By dimensional analysis, one has
\begin{equation}
p=n^{5/3}{\cal F}(t,c), \quad t=\frac{T}{n^{2/3}}, \quad c=a n^{1/3},
\end{equation}
with ${\cal F}(0,c)=4\pi c$ (in Bogoliubov theory), and ${\cal F}(t,0)=\frac{\zeta(5/2)}{(4\pi)^{3/2}}t^{5/2}$ (free theory), such that for small $c$
\begin{equation}
{\cal F}=\frac{\zeta(5/2)}{(4\pi)^{3/2}}t^{5/2}+g(t)c.
\end{equation}

At zero temperature the second term in eq. \eqref{eqSoundAdiabaticIsothermal} vanishes, such that $v_S=v_T$. For the isothermal sound velocity one has 
\begin{equation}
v_T^2=2\frac{\partial p}{\partial n}\bigg{|}_{T}=2 \frac{\partial p}{\partial \sigma}\bigg{|}_T\left(\frac{\partial n}{\partial \sigma}\bigg{|}_T\right)^{-1}.
\end{equation}
We can now use 
\begin{equation}
\frac{\partial p}{\partial \sigma}\big{|}_T=-\frac{dU_{\text{min}}}{d\sigma}=-\partial_\sigma U(\rho_0)=n
\end{equation}
and infer
\begin{equation}
v_T^2=2\left(\frac{\partial\, \text{ln}\,n}{\partial \sigma}\right)^{-1}.
\end{equation}

One may also define a microscopic sound velocity $c_S$, which characterizes the propagation of (quasi-) particles. At zero temperature, where we can perform the analytic continuation to real time, we can calculate the microscopic sound velocity from the dispersion relation $\omega(p)$ (with $p=|\vec{p}|$). In turn, the dispersion relation is obtained from the effective action by setting $\text{det}(G^{-1})=0$, where $G^{-1}$ is the full inverse propagator. We perform the calculation explicitly in app. \ref{sectPropDisp} and find
\begin{equation}
c_S^{-2}=\frac{S^2}{2\lambda\rho_0}+V.
\end{equation}

\begin{figure}
\includegraphics{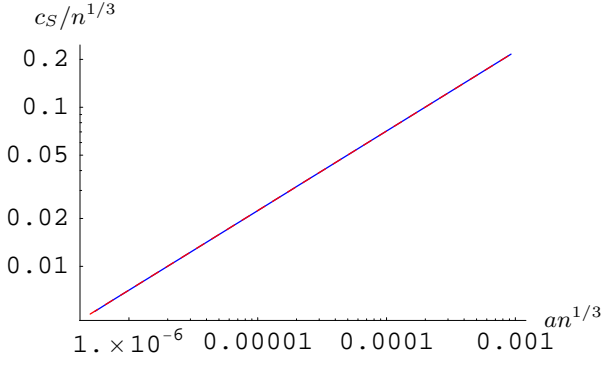}
\caption{(Color online) Dimensionless sound velocity $c_s/(n^{1/3})$ at zero temperature, as a function of the scattering length $an^{1/3}$. Within the plot resolution the curves obtained by varying $a$ with fixed $n$, by varying $n$ with fixed $a$, and the Bogoliubov result, $c_s=\sqrt{16\pi}(an)^{1/2}$, coincide.}
\label{figSound}
\end{figure}

The Bogoliubov result for the sound velocity is in our units
\begin{equation}
c_S^2=2\lambda\rho_0=16\pi an.
\end{equation}
In three dimensions, the decrease of $S$ is very slow and the coupling $V$ remains comparatively small even on macroscopic scales, cf. fig. \ref{figFlowKinetic}. We thus do not expect measurable deviations from the Bogoliubov result for the sound velocity at $T=0$. In figure \ref{figSound}, we plot our result over several orders of magnitude of the dimensionless scattering length and, indeed, find no deviations from Bogoliubovs result.

We finally show that for $T=0$ the macroscopic and microscopic sound velocities are equal, $v_S=v_T=c_S$. For this purpose, we use 
\begin{equation}
\frac{\partial n}{\partial \sigma}\big{|}_T=-\frac{d}{d\sigma}\left(\partial_\sigma U(\rho_0)\right)=-\partial_\sigma^2U(\rho_0)-\partial_\rho\partial_\sigma U(\rho_0)\frac{d\rho_0}{d\sigma}.
\end{equation}
From the minimum condition $\partial_\rho U=0$, it follows
\begin{equation}
\frac{d\rho_0}{d\sigma}=-\frac{\partial_\rho\partial_\sigma U}{\partial_\rho^2 U}=-\frac{\alpha}{\lambda}.
\end{equation}
Combining this with the Ward identities from app. \ref{sectTruncation}, namely $\partial_\sigma^2 U=-2V\rho_0$ and $\alpha=\partial_\rho\partial_\sigma U=-S$, valid at $T=0$, it follows that the macroscopic sound velocity equals the microscopic sound velocity
\begin{equation}
v_S^2(T=0)=c_S^2.
\end{equation}

\section{Conclusions}
\label{sectConclusions}
The use of functional renormalization yields a quantitative description of the whole phase diagram for dilute non-relativistic bosons in three dimensions. This can describe a gas of ultracold bosonic atoms. More generally, our results can also be applied to quantum phase transitions or low temperature phase transitions, whenever the most relevant excitations correspond to bosonic quasi-particles. For this reason we deal with a general dispersion relation, involving in the classical propagator terms linear and quadratic in the frequency. As a function of temperature $T$ and effective chemical potential $\sigma$, we have computed the pressure $p$, the density $n$, condensate $\bar{\rho}_0$, superfluid density $\rho_0$, and the sound velocity $c_S$, in dependence on two system parameters, namely the scattering length $a$ and a dimensionless coupling $\tilde{v}$ parameterizing the classical dispersion relation. For $T=0$ and $\tilde{v}=0$ we find very good agreement with the Bogoliubov theory. As $T$  increases, the condensate melts at a critical temperature $T_c$, which exceeds the one for the free theory by $\Delta  T_c/T_c=2.1 an^{1/3}$. We find a second order phase transition in the universality class of the three-dimensional $O(2)$ model, with the associated universal critical exponents.

We have found an upper bound for the scattering length, which is of the order of the microscopic length scale $\Lambda^{-1}$. This indicates the breakdown of the pointlike approximation. For atom gases, $\Lambda^{-1}$ typically corresponds to the range of the Van der Waals interaction. The upper bound is a result of strong fluctuation effects in presence of a large microscopic interaction. Even for an arbitrary strong microscopic interaction, the quantum fluctuations reduce the renormalized coupling, which corresponds to the physical scattering length in vacuum. For non-relativistic atoms with pointlike interactions, a strong interaction regime can therefore only be realized at high density, $n\sim \Lambda^3$. Superfluid liquids, like $^4\text{He}$ are good examples for such systems. First functional renormalization studies for such systems have already been performed \cite{Gollisch:2001ff}. Our extended treatment overcomes several problems of this early approach and we will apply it to $^4\text{He}$ in the future. Except for the vicinity of the phase transition, however, the detailed microphysics may play an important role, since all relevant length scales are of the order of the microphysical length scale $\Lambda^{-1}$. Other interesting extensions concern lower dimensional systems. We look forward to a unified functional renormalization description of bosons with pointlike interactions, for arbitrary dimensions and arbitrary interaction strength.

\begin{appendix}
\section{Derivative expansion and Ward identities\label{sectTruncation}}
We use a derivative expansion for the truncation of the effective average action with derivative operators up to four momentum dimensions 
\begin{eqnarray}
\nonumber
\Gamma_k &=& \int_x{\Bigg \{}U(\rho,\sigma)\\
\nonumber
&&+\frac{1}{2}Z_1(\rho,\sigma) \left(\phi^*\partial_\tau\phi-\phi\partial_\tau\phi^*\right)\\
\nonumber
&&+\frac{1}{2}Z_2(\rho,\sigma) \left(\phi^*(-\Delta)\phi+\phi(-\Delta)\phi^*\right)\\
\nonumber
&&+\frac{1}{2}V_1(\rho,\sigma) \left(\phi^*(-\partial_\tau^2)\phi+\phi(-\partial_\tau^2)\phi^*\right)\\
\nonumber
&&+V_2(\rho,\sigma) \left(\phi^*(\partial_\tau\Delta)\phi-\phi(\partial_\tau\Delta)\phi^*\right)\\
&&+\frac{1}{2}V_3(\rho,\sigma) \left(\phi^*(-\Delta^2)\phi+\phi(-\Delta^2)\phi^*\right){\Bigg \}}
\label{derivativeexpansion}
\end{eqnarray}
Here, we employ the renormalized fields 
\begin{eqnarray}
\nonumber
\phi &=& \bar{A}^{1/2}\bar{\phi},\\
\rho &=& \phi^*\phi=\bar{A}\bar{\rho}=\bar{A}\bar{\phi}^*\bar{\phi}
\end{eqnarray}
and coupling functions $U$, $Z_i$, $V_i$. We fix the wave function renormalization factor $\bar{A}$ such that $Z_1(\rho_0,\sigma_0)=1$. 
Terms of the form $\rho(-\Delta)\rho$ or $\rho(-\partial_\tau^2)\rho$ are not included here, since they are expected to play a sub-leading role. For a systematic derivative expansion they have to be added - the terms with up to two derivatives can be found in \cite{Wetterich:2007ba}. In terms of dimensions, the operator $\partial_\tau$ counts as two space derivatives for the nonrelativistic model with $V=0$, while it counts as one space dimension for the relativistic model with $S=0$. We expand the $k$-dependent functions $U(\rho,\sigma)$, $Z_1(\rho,\sigma)$, $Z_2(\rho,\sigma)$, $V_1(\rho,\sigma)$,$V_2(\rho,\sigma)$ and $V_3(\rho,\sigma)$ around the $k$-dependent minimum $\rho_0(k)$ of the effective potential and the $k$-independent value of the chemical potential $\sigma_0$ that corresponds to the physical particle number density $n$. For example, with $Z_1=Z_1(\rho_0,\sigma_0)$, one has
\begin{eqnarray}
\nonumber
Z_1(\rho,\sigma)&=&Z_1+Z_1^\prime(\rho_0,\sigma_0)(\rho-\rho_0)\\
&&+Z_1^{(\sigma)}(\rho_0,\sigma_0)(\sigma-\sigma_0).
\end{eqnarray}

Let us concentrate on the non-relativistic model where $S=1$, $V=0$ in the microscopic action. At zero temperature, we can perform an analytic continuation to real time $\tau=it$. The microscopic action \eqref{microscopicaction} is then
\begin{eqnarray}
\nonumber
S[\phi]&=&-\int_{-\infty}^{\infty} dt\int d^3x \\
&&{\Big \{}\phi^*\,(-i\partial_t-\sigma-\Delta)\,\phi\,+\,\frac{1}{2}\lambda(\phi^*\phi){\Big \}}.
\label{realtimemicroscopicaction}
\end{eqnarray}
In addition to space translations, rotations and time translations, two further symmetries constrain the possible forms of the couplings in $\Gamma$. In order to derive these constraints, we extend eq. \eqref{realtimemicroscopicaction} to a $t$-dependent source $\sigma(t)$. First, there is a semi-local $U(1)$ symmetry of the form
\begin{eqnarray}
\nonumber
\phi(t,\vec{x})& \rightarrow & e^{i\alpha(t)}\phi(t,\vec{x})\\
\nonumber
\phi^*(t,\vec{x}) & \rightarrow & e^{-i\alpha(t)}\phi^*(t,\vec{x})\\
\sigma & \rightarrow  & \sigma+\partial_t\alpha.
\end{eqnarray}
This holds since the combination $(-i\partial_t-\sigma)$ acts as a covariant derivative. In addition, we have the invariance under Galilean boost transformations of the fields
\begin{eqnarray}
\nonumber
\phi(t,\vec{x}) & \rightarrow & \phi'(t,\vec{x})=e^{-i(\vec{q}^2t-\vec{q}\vec{x})}\phi(t,\vec{x}-2\vec{q}t)\\
\phi^*(t,\vec{x}) & \rightarrow & \phi^{*\prime}(t,\vec{x})=e^{i(\vec{q}^2t-\vec{q}\vec{x})}\phi^*(t,\vec{x}-2\vec{q}t).\label{galileanboost}
\end{eqnarray}
While the invariance of the interaction term under this symmetry is obvious, its realization for the kinetic term is more involved. Performing the transformation explicitly, one finds
\begin{eqnarray}
\nonumber
\phi^*\Delta\phi & \rightarrow & \phi^*\Delta\phi-\vec{q}^2\phi^*\phi+2 i\vec{q}\phi^*\vec{\nabla}\phi\\
\phi^*i\partial_t\phi & \rightarrow & \phi^*i\partial_t\phi+\vec{q}^2\phi^*\phi-2 i\vec{q}\phi^*\vec{\nabla}\phi,
\end{eqnarray}
such that indeed the combination
\begin{equation}
i\partial_t+\Delta \label{schroedingeroperator}
\end{equation}
leads to this invariance. On the other hand, the validity of the Galilean symmetry for an effective action guarantees that only the combination \eqref{schroedingeroperator} or powers of this operator act on $\phi$. An operator of the form $(i\partial_t+\gamma\Delta)$ with $\gamma\neq 1$ would break the symmetry. (Note, that $\Delta\rho$ is also invariant.)

Both the semi-local $U(1)$ symmetry and the Galilean symmetry are helpful only at zero temperature. At nonzero temperature, the analytic continuation to real time is no longer useful. An analog version of the semi-local $U(1)$ transformation for Euclidean time $\tau$ would involve the imaginary part of the chemical potential $\sigma$, which has no physical meaning. The dependence of physical quantities on $\sigma+\sigma^*$ is not restricted. In addition, the Galilean symmetry is broken explicitly by the thermal heat bath.
 
Combining semi-local $U(1)$ symmetry and Galilean symmetry at $T=0$, we find that the derivative operators $i\partial_t$, $\Delta$ and the chemical potential term $(\sigma-\sigma_0)$ are combined to powers of the operator
\begin{equation}
D=(-i\partial_t-(\sigma-\sigma_0)-\Delta).
\end{equation}
In addition to powers of that operator acting on $\phi$, only spatial derivatives of terms, that are invariant under $U(1)$ transformations, like $\rho\Delta\rho$, may appear. Since the symmetry transformations act linearly on the fields, the full effective action $\Gamma[\phi]$ is also invariant. This also holds for the average action $\Gamma_k[\phi]$, provided that the cutoff term $\Delta S_k[\phi]$ is invariant. We can write the effective action as an expansion in the operator $D$
\begin{eqnarray}
\nonumber
\Gamma[\phi]&=&\int_x \bigg{\{}U_0(\rho)\\
\nonumber
&&+\frac{1}{2}\tilde{Z}(\rho)\left(\phi^*(-i\partial_t-(\sigma-\sigma_0)-\Delta)\phi+c.c\right)\\
\nonumber
&&+\frac{1}{2}\tilde{V}(\rho)\left(\phi^*(-i\partial_t-(\sigma-\sigma_0)-\Delta)^2\phi+c.c\right)\\
&&+\dots\bigg{\}}
\label{eqEffectiveactionGalilean}
\end{eqnarray}
Performing the Wick rotation back to Euclidean time, we can compare this to eq. \eqref{derivativeexpansion}, and find for $T=0$ the relations
\begin{eqnarray}
\nonumber
Z_1(\rho,\sigma_0)&=&Z_2(\rho,\sigma_0)=\tilde{Z}(\rho),\\
\nonumber
V_1(\rho,\sigma_0)&=&V_2(\rho,\sigma_0)=V_3(\rho,\sigma_0)=\tilde{V}(\rho),\\
\nonumber
Z_1^{(\sigma)}(\rho_0,\sigma_0)&=&2\left(\tilde{V}(\rho_0)+\rho_0\tilde{V}^\prime(\rho_0)\right),\\
Z_2^{(\sigma)}(\rho_0,\sigma_0)&=&2\tilde{V}(\rho_0),
\end{eqnarray}
and therefore
\begin{eqnarray}
\nonumber
\alpha&=&-\left(\tilde{Z}(\rho_0)+\rho_0\tilde{Z}^\prime(\rho_0)\right),\\
\nonumber
n_k&=&\tilde{Z}(\rho_0)\rho_0,\\
\beta &=&-\left(2\tilde{Z}^\prime(\rho_0)+\rho_0\tilde{Z}^{\prime\prime}(\rho_0)\right).
\end{eqnarray}

We next compute the inverse propagator in a constant background field by expanding $\Gamma_k$ to second order in the fluctuations around this background. For this purpose, it is convenient to decompose 
\begin{equation}
\phi(\tau,\vec{x})=\phi_0+\frac{1}{\sqrt{2}}\left(\phi_1(\tau, \vec{x})+i\phi_2(\tau,\vec{x})\right).
\end{equation}
The constant condensate field $\phi_0$ can be chosen to be real without loss of generality. The fluctuating real fields are the radial mode $\phi_1$ and the Goldstone mode $\phi_2$, and $\rho=\rho_0+\sqrt{2}\phi_0\phi_1+\frac{1}{2}\phi_1^2+\frac{1}{2}\phi_2^2$. The truncation of the effective average action \eqref{derivativeexpansion} reads in that basis
\begin{widetext}
\begin{eqnarray}
\nonumber
\Gamma_k[\phi] = \int_x{\Bigg \{}U(\rho,\sigma)
%\\
%\nonumber
%&&
&+& \frac{1}{2}Z_1(\rho,\sigma) \left(i\sqrt{2}\phi_0\partial_\tau\phi_2+i\phi_1\partial_\tau \phi_2-i\phi_2\partial_\tau\phi_1\right)\\
\nonumber
&+&\frac{1}{2}Z_2(\rho,\sigma) \left(\sqrt{2}\phi_0(-\Delta)\phi_1+\phi_1(-\Delta) \phi_1+\phi_2(-\Delta)\phi_2\right)\\
\nonumber
&+&\frac{1}{2}V_1(\rho,\sigma) \left(\sqrt{2}\phi_0(-\partial_\tau^2)\phi_1+\phi_1(-\partial_\tau^2) \phi_1+\phi_2(-\partial_\tau^2)\phi_2\right)\\
\nonumber
&+&V_2(\rho,\sigma) \left(i\sqrt{2}\phi_0(\partial_\tau\Delta)\phi_2+i\phi_1(\partial_\tau\Delta) \phi_2-i\phi_2(\partial_\tau\Delta)\phi_1\right)\\
&+&\frac{1}{2}V_3(\rho,\sigma) \left(\sqrt{2}\phi_0(-\Delta^2)\phi_1+\phi_1(-\Delta^2) \phi_1+\phi_2(-\Delta^2)\phi_2\right){\Bigg \}}.
\end{eqnarray}
The inverse propagator $\Gamma_k^{(2)}$ can be inferred from an expansion in second order in $\phi_1$ and $\phi_2$. We keep the linear order in $\sigma-\sigma_0$, which will be needed for the flow equation for the density. This yields
\begin{eqnarray}
\nonumber
\Gamma_k[\phi] &=&\int_x{\Bigg \{}U(\rho_0,\sigma_0)+U^{(\sigma)}\,(\sigma-\sigma_0)+\frac{1}{2}(U^\prime+2\rho_0 U^{\prime\prime})\phi_1^2+\frac{1}{2}U^\prime \phi_2^2\\
\nonumber
&+&\frac{1}{2}\left(Z_1+Z_1^\prime \rho_0+Z_1^{(\sigma)}(\sigma-\sigma_0)\right) \left(i\phi_1\partial_\tau \phi_2-i\phi_2\partial_\tau\phi_1\right)\\
\nonumber
&+&\frac{1}{2}\left(1+2 Z_2^\prime \rho_0+Z_2^{(\sigma)}(\sigma-\sigma_0)\right) \left(\phi_1(-\Delta) \phi_1\right)\\
\nonumber
&+&\frac{1}{2}\left(1+Z_2^{(\sigma)}(\sigma-\sigma_0)\right) \left(\phi_2(-\Delta) \phi_2\right)\\
\nonumber
&+&\frac{1}{2}\left(V_1+2 V_1^\prime\rho_0+V_1^{(\sigma)}(\sigma-\sigma_0)\right) \left(\phi_1(-\partial_\tau^2) \phi_1\right)\\
\nonumber
&+&\frac{1}{2}\left(V_1+V_1^{(\sigma)}(\sigma-\sigma_0)\right) \left(\phi_2(-\partial_\tau^2) \phi_2\right)\\
\nonumber
&+&\left(V_2+V_2^\prime\rho_0+V_2^{(\sigma)}(\sigma-\sigma_0)\right) \left(i\phi_1(\partial_\tau\Delta) \phi_2-i\phi_2(\partial_\tau\Delta)\phi_1\right)\\
\nonumber
&+&\frac{1}{2}\left(V_3+2V_3^\prime \rho_0+V_3^{(\sigma)}(\sigma-\sigma_0)\right) \left(\phi_1(-\Delta^2) \phi_1\right)\\
&+&\frac{1}{2}\left(V_3+V_3^{(\sigma)}(\sigma-\sigma_0)\right)\left(\phi_2(-\Delta^2) \phi_2\right){\Bigg \}},
\end{eqnarray}
\end{widetext}
where we dropped the argument $(\rho_0,\sigma_0)$ at several places and used the implicit rescaling condition $Z_2(\rho_0,\sigma_0)=1$.
In our simple truncation, we take at $\sigma=\sigma_0$ only 
\begin{eqnarray}
\nonumber
S&=&Z_1(\rho_0,\sigma_0)+Z_1^\prime(\rho_0,\sigma_0) \,\rho_0,\\
V&=&V_1(\rho_0,\sigma_0)
\end{eqnarray} 
into account. We neglect the contribution of the other couplings, i.e. set $Z_2^\prime=V_2=V_3=V_1^\prime=V_2^\prime=V_3^\prime=0$. As shown above, it follows from the symmetry requirements at zero temperature, that  $V_1=V_2=V_3=\tilde{V}$, $Z_1^{(\sigma)}=2(\tilde{V}+\tilde{V}^\prime\rho_0)$ and $Z_2^{(\sigma)}=2\tilde{V}$. The truncation $V_2=V_3=0$ therefore violates the Galilean symmetry, as does our choice of the cutoff term $\sim R_k$. Within our approximation, it is consistent to set $Z_1^{(\sigma)}=Z_2^{(\sigma)}=2V$ at zero temperature. Also the deviations from this relation at finite temperature are neglected for simplicity in this paper. This yields eq. \eqref{eqSimpleTruncation}.

\section{Symmetries and Noether currents}
\label{sectSymmetriesandNoethercurrents}
In the following we discuss the role of continuous symmetries of the microscopic action $S[\phi]$. Since all these symmetries are linear in the fields, the full effective action $\Gamma[\phi]$ is also symmetric. In the case that the cutoff term $\Delta S_k[\phi]$ is chosen invariant under the symmetry transformation in question, this also holds for the average action $\Gamma_k[\Phi]$ for finite $k$. From Noether's theorem it follows that there exists a conserved current $j^\mu=(j^0,\vec{j})$ connected with every such symmetry. If the action is formulated as an integral over the imaginary time $\tau$ the conservation equation implies for the current
\begin{equation}
\partial_\tau j^{(\tau)}+\vec{\nabla}\vec{j}=0. \label{euklidconservedcurrent}
\end{equation}
At zero temperature, we can perform a Wick rotation to real time, $\tau\rightarrow it$, and eq. \eqref{euklidconservedcurrent} takes the usual form
\begin{equation}
\partial_t j^{(t)}+\vec{\nabla}\vec{j}=0. \label{realtimeconservedcurrent}
\end{equation}
The Noether charge $C=\int d^3 x j^{(t)}$ is conserved in time, i.e. $\frac{d}{dt}C=0$. This holds if $\vec{j}$ falls off sufficiently fast at spatial infinity. At finite temperature however, the situation is different. A simple analytic continuation to real time is no longer possible, since the configuration space is now a torus with periodicity $1/T$ in the $\tau$-direction. Instead, we can integrate eq.  \eqref{euklidconservedcurrent} over complex time $\tau$, giving
\begin{equation}
\vec{\nabla}\vec{J}=\vec{\nabla}\int_0^{1/T}d\tau \vec{j}=j^{(\tau)}(0)-j^{(\tau)}(1/T)=0.
\end{equation}
From the symmetry, it now follows that there exists a solenoidal vector field or three component current $\vec{J}=\int_{\tau}\vec{j}$.

A global symmetry of an action $\Gamma[\phi]$ (where $\Gamma$ could be replaced by $S$ or $\Gamma_k$ if appropriate) can be formulated in its infinitesimal form as
\begin{equation}
\Gamma[\phi+\epsilon s\phi]=\Gamma[\phi],
\end{equation}
with $\epsilon$ independent of $x$. Here $s$ is the infinitesimal generator of the symmetry transformation. For a local transformation, where $\epsilon$ depends on $x$, $\epsilon=\epsilon(x)$, we can expand
\begin{equation}
\Gamma [\phi+\epsilon s \phi]=\Gamma[\phi]+\int_x \left\{ (\partial_\mu\epsilon){\cal J}^\mu+(\partial_\mu\partial_\nu\epsilon){\cal K}^{\mu\nu}+\dots\right\}.
\label{Noetherexpansion}
\end{equation}
The global symmetry implies that the expansion on the r.h.s. of eq. \eqref{Noetherexpansion} starts with $\partial_\mu\epsilon$. Here and in the following it is implied that $\epsilon$ as well as its derivatives are infinitesimal, i.e. we keep only terms that are linear in $\epsilon$. The index $\mu$ goes over $(0,1,2,3)$, representing $(t,x^1,x^2,x^3)$ in the real time case and $(\tau,x^1,x^2,x^3)$ for imaginary time. Eq. \eqref{Noetherexpansion} implies for arbitrary $\phi(x)$
\begin{equation}
\int_x\left\{\frac{\delta\Gamma[\phi]}{\delta\phi}\epsilon s\phi-(\partial_\mu\epsilon){\cal J}^\mu-(\partial_\mu\partial_\nu\epsilon){\cal K}^{\mu\nu}+\dots\right\}=0.
\label{eqNoetherIntegral}
\end{equation}
Our notation is for real fields and implies a summation over components, if appropriate. In a complex basis one replaces $\frac{\delta \Gamma}{\delta \phi}\epsilon s\phi$ by $\frac{\delta \Gamma}{\delta \phi}\epsilon s\phi+\frac{\delta \Gamma}{\delta \phi^*}\epsilon s\phi^*$.

Eq. \eqref{eqNoetherIntegral} is valid for all field configurations $\phi$ and not only for those that fulfill the field equation $\frac{\delta\Gamma[\phi]}{\delta \phi}=0$. In consequence, the integrand is a total derivative
\begin{eqnarray}
\nonumber
& \frac{\delta\Gamma[\phi]}{\delta\phi}\epsilon s\phi-(\partial_\mu\epsilon){\cal J}^\mu-(\partial_\mu\partial_\nu\epsilon){\cal K}^{\mu\nu}+\ldots &\\
& =-\partial_\mu\left(j^\mu \epsilon+\kappa^{\mu\nu}\partial_\nu\epsilon+\ldots\right). &
\end{eqnarray}
We can now specialize to $\partial_\mu\epsilon=\partial_{\mu}\partial_\nu\epsilon=\ldots=0$ and find
\begin{equation}
\frac{\delta\Gamma[\phi]}{\delta\phi} s\phi(x)=-\partial_\mu j^\mu.
\end{equation}
This defines the Noether current $j^\mu$. For solutions of the field equation, $\frac{\delta\Gamma[\phi]}{\delta\phi}=0$, the current $j^\mu$ is conserved, $\partial_\mu j^\mu=0$.

For a given $x$ we can also specialize to
\begin{equation}
\epsilon(x)=0, \,\,\,\partial_\mu\epsilon(x)\neq0, \,\,\,\partial_\mu\partial_\nu\epsilon(x)=0,\,\,\,\ldots,
\end{equation}
which leads to 
\begin{equation}
j^\mu={\cal J}^\mu-\partial_\nu \kappa^{\nu\mu}.
\label{currentcorrection}
\end{equation}
This process can be continued, leading us to a whole tower of identities for the conserved current $j^\mu$. 

If the action $\Gamma[\phi]$ includes derivatives only up to a finite order $n$, i.e. can be written in the form
\begin{equation}
\Gamma[\phi]=\int_x{\cal L}(\phi,\partial\phi, \partial\partial\phi, \ldots, \partial^{(n)}\phi),
\end{equation}
the expansion on the right hand side of \eqref{Noetherexpansion} only contains terms up to order $\partial^{(n)}\epsilon$ such that the tower of equations for $j^\mu$ can be solved.
Moreover, for homogeneous situations where $\frac{\delta\Gamma[\phi]}{\delta\phi}$ is solved by a constant $\phi$, the second term on the r.h.s. of eq. \eqref{currentcorrection} vanishes since it includes a derivative. We have then $j^\mu={\cal J}^\mu$.

A convenient way to find the local currents employs parameters $\epsilon(x)$ that decay sufficiently fast at infinity such that we can partially integrate eq. \eqref{eqNoetherIntegral}
\begin{equation}
\int_x\epsilon(x)\left\{\frac{\delta\Gamma[\phi]}{\delta\phi} s\phi+\partial_\mu{\cal J}^\mu-\partial_\mu\partial_\nu{\cal K}^{\mu\nu}+\dots\right\}=0.
\end{equation}
This yields the local identity
\begin{equation}
\frac{\delta \Gamma[\phi]}{\delta \phi}s\phi=-\partial_\mu{\cal J}^\mu+\partial_\mu\partial_\nu {\cal K}^{\mu\nu}-\dots.
\end{equation}
An expansion of the l.h.s. in derivatives often yields substantial information on ${\cal J}^\mu$ etc. by inspection. 

Our construction yields a unique conserved local current $j^\mu$ for every generator of a continuous symmetry. We note, however, that $\alpha j^\mu+b^\mu$ is also a conserved local current if $\alpha$ and $b^\mu$ are independent of $x$. This remark is important if we want to associate $j^\mu$ with the current for a physical quantity. A rotation invariant setting implies $b^i=0$, but $b^0$ and $\alpha$ may differ from zero. 

After these general considerations we now specialize to nonrelativistic real time actions of the form
\begin{equation}
\Gamma[\phi]=\int_{-\infty}^{\infty}dt \int d^3x{\cal L}(\phi, (i\partial_t+\Delta)\phi,(i\partial_t+\Delta)^2\phi,\dots).
\end{equation}
We assume, that $\Gamma$ invariant under the same symmetries as the action \eqref{realtimemicroscopicaction}. From the symmetry under time translations
\begin{eqnarray}
\nonumber
\phi & \rightarrow & \phi+\epsilon (s_t)\phi=\phi+\epsilon \partial_t \phi\\
{\cal L}& \rightarrow & {\cal L}+\epsilon \partial_t{\cal L}={\cal L}+\epsilon\partial_\mu(\delta^\mu_0 {\cal L}),
\end{eqnarray}
we find a conserved current $(j_E)^\mu$. Up to a possible additive constant its $t$-component is the energy density, while the spatial components describe energy flux density. The multiplicative constant $\alpha$ gets fixed if we choose the units to measure energy. The choice $\hbar=1$ corresponds to $\alpha=1$. Similarly, the invariance under spatial translations
\begin{eqnarray}
\nonumber
\phi & \rightarrow  & \phi+\epsilon^i(s_M)_i \phi=\phi-\epsilon^i\partial_i\phi\\
{\cal L} & \rightarrow & {\cal L}-\epsilon^i\partial_i{\cal L}={\cal L}-\epsilon^i\partial_\mu(\delta_i^\mu{\cal L})
\end{eqnarray}
implies a conserved current $(j_M)_i^\mu$ for each spatial direction $i=1,2,3$. Up to an additive constant $(b_M)^0_i$ the $t$-component is the conserved momentum density, $p_i=(j_{M})^0_i+(b_M)^0_i$, while the spatial components can be interpreted as a momentum flux density, with the diagonal components $(j_{M})^i_i$ describing pressure. 

From the global $U(1)$ symmetry
\begin{eqnarray}
\nonumber
\phi & \rightarrow & \phi+\epsilon (s_C)\phi=\phi-i \epsilon \phi\\
\nonumber
\phi^* & \rightarrow & \phi^*+\epsilon (s_C) \phi^*=\phi^*+i\epsilon \phi^*\\
{\cal L} & \rightarrow & {\cal L}
\end{eqnarray}
we can infer the conservation of the current $(j_C)^\mu$
associated to the conserved particle number.  In order to identify the total particle number with the charge of this current, $\int d^3x (j_C)^0$, we need to fix a possible multiplicative constant $\alpha$. For this purpose, we use the Galilean boost invariance, described already in eq. \eqref{galileanboost}. It reads in its infinitesimal form
\begin{eqnarray}
\nonumber
\phi & \rightarrow & \phi+\epsilon^i(s_G)_i\phi=\phi+2\epsilon^it\partial_i\phi-i\epsilon^i x_i\phi\\
\nonumber
\phi^* & \rightarrow & \phi^*+\epsilon^i(s_G)_i \phi^* = \phi^*+2\epsilon^i t\partial_i\phi^*+i\epsilon^i x_i\phi^*\\
{\cal L} & \rightarrow & {\cal L}+\epsilon^i\partial_\mu(2\,\delta^\mu_i t{\cal L}),
\end{eqnarray}
and the conserved charge of $(s_G)$ is the center of mass, again up to an additive constant. The generator $(s_G)$ can be decomposed as
\begin{equation}
(s_G)_i=x_i(s_C)-2t(s_M)_i. 
\end{equation}
This implies for the current
\begin{equation}
(j_G)^\mu_i=x_i\,(j_C)^\mu-2t\,(j_M)^\mu_i.
\end{equation}
Specializing to the $t$-component, identifying the momentum density $p_i=(j_M)^0_i+(b_M)^0_i$ and reintroducing the particle mass $2M=1$ we find
\begin{equation}
(j_G)^0_i=x_i\,(j_C)^0-t\frac{p_i-(b_M)^0_i}{M}. 
\end{equation}
From this we can conclude that up to an additive constant $(j_C)^0$ is the particle density $n=(j_C)^0+(b_C)^0$. 

For the effective action \eqref{eqEffectiveactionGalilean} we find for $\sigma=\sigma_0$ and constant $\phi(x)=\sqrt{\rho_0}$ the current
\begin{equation}
(j_C)^0=\tilde{Z}(\rho_0)\rho_0.
\end{equation}
Using the normalization condition $\tilde{Z}(\rho_0)=1$, this gives $(j_C)^0=\rho_0$. At zero temperature, this is the particle density and the additive constant $(b_C)^0$ vanishes. At nonzero temperature we can compare to eq. \eqref{eqDensityatT} and find $(b_C)^0=n_T$.

For completeness we also mention the symmetry under spatial rotations
\begin{eqnarray}
\nonumber
\phi(t,\vec{x})\rightarrow\phi(t,R^{-1}\vec{x})\\
{\cal L}(t,\vec{x})\rightarrow{\cal L}(t,R^{-1}\vec{x}),
\end{eqnarray}
with orthogonal matrix $R^i_{\,j}=(e^{i\vec{\eta}\vec{J}})^i_{\,j}$, generators $(J_i)^j_{\,k}=i\varepsilon_{ijk}$, and $\varepsilon_{ijk}$ the antisymmetric tensor in three dimensions.
The infinitesimal transformation reads
\begin{eqnarray}
\nonumber
\phi(t,\vec{x})\rightarrow\phi(t,\vec{x})+\eta^i\varepsilon_{ijk}x^k\partial_j\phi(t,\vec{x})\\
{\cal L}(t,\vec{x})\rightarrow{\cal L}(t,\vec{x})+\eta^i\partial_l(\varepsilon_{ijk}x^k\delta^l_j{\cal L}).
\end{eqnarray}
The time component of the conserved current $(j_R)_i^\mu$ is, of course, the angular momentum density. 

\section{Flow equation for effective potential}
\label{secFlowEffectivePotential}
We derive the flow equation for the effective potential by evaluating the flow equation for the average action \eqref{eqFlowequation} for constant fields. Inserting a real constant field $\phi(x)=\sqrt{\rho}$ one finds for $U=\Gamma_k/\Omega$ the flow at fixed $\rho$
\begin{eqnarray}
\nonumber
&& \partial_t U(\rho,\sigma) \,=\, \eta \rho U^\prime\,+ \,\zeta(\rho,\sigma),\\
\nonumber
&& \zeta(\rho,\sigma)=T\sum_n 2 v_d \int_0^\infty dp\, p^{d-1}\theta(k^2-p^2-m^2)\\
\nonumber
&& \left(2k^2-\eta(k^2-p^2-m^2)+\partial_t m^2\right)\\
&& \frac{g_1+g_2+2(V_1+\rho V_1^\prime)\omega_n^2}{h^2 \omega_n^2+(g_1+(V_1+2\rho V_1^\prime)\omega_n^2)(g_2+V_1 \omega_n^2)}.
\end{eqnarray}
Here, $d$ is the number of spatial dimensions and we use the abbreviations
\begin{eqnarray}
\nonumber
g_1 &=& k^2-m^2+(Z_2-1+2\rho Z_2^\prime)p^2-(V_3+2\rho V_3^\prime)p^4\\
\nonumber
&&+U^\prime+2\rho U^{\prime\prime},\\
\nonumber
g_2 &=& k^2-m^2+(Z_2-1)p^2-V_3p^4+U^\prime,\\
\nonumber
h &=& Z_1+\rho Z_1^\prime -(V_2+\rho V_2^\prime)p^2,\\
\nonumber
\omega_n&=&2\pi T n,\\
v_d &=& (2^{d+1}\pi^{d/2}\Gamma(d/2))^{-1}.
\label{eqabbeffpot}
\end{eqnarray}
We dropped the arguments $(\rho,\sigma)$ at several places on the right hand side. Primes denote derivatives with respect to $\rho$. In the phase with spontaneous symmetry breaking, we have $m^2=\partial_tm^2=0$. 

The Matsubara sums over $n$ can be carried out by virtue of the formulae
\begin{eqnarray}
\nonumber
&& \sum_{n=-\infty}^{\infty}\frac{1}{an^4+bn^2+c}=\frac{\pi}{d\sqrt{2c}} \\
\nonumber
&& \left(\sqrt{b+d}\,\text{coth}(\sqrt{\frac{b-d}{2a}}\pi)-\sqrt{b-d}\,\text{coth}(\sqrt{\frac{b+d}{2a}}\pi)\right), \\
\nonumber
&& \sum_{n=-\infty}^{\infty}\frac{n^2}{an^4+bn^2+c}=\frac{\pi}{d\sqrt{2a}} \\
\nonumber
&& \left(\sqrt{b+d}\,\text{coth}(\sqrt{\frac{b+d}{2a}}\pi)-\sqrt{b-d}\,\text{coth}(\sqrt{\frac{b-d}{2a}}\pi)\right),\\
\label{eqMatsubara}
\end{eqnarray}
with $d=\sqrt{b^2-4ac}$. This brings us to
\begin{eqnarray}
\nonumber
&& \zeta(\rho,\sigma)= 2 v_d\int_0^{\sqrt{k^2-m^2}}dp\,p^{d-1}\\
\nonumber
&& (2k^2-\eta(k^2-p^2-m^2)+\partial_tm^2)\frac{1}{\sqrt{8}D}\\
\nonumber
&& \Bigg{(}\left(\sqrt{B+D}\frac{E}{\sqrt{C}}-2\sqrt{B-D}\right)\,\text{coth}(\frac{\sqrt{B-D}}{\sqrt{8A}\,T})\\
\nonumber
&& + \left(2\sqrt{B+D}-\sqrt{B-D}\frac{E}{\sqrt{C}}\right)\,\text{coth}(\frac{\sqrt{B+D}}{\sqrt{8 A}\,T})\Bigg{)},\\
\label{eqFlowEffectivPotentialSystematic}
\end{eqnarray}
where we introduced 
\begin{eqnarray}
\nonumber
A & = & V_1(V_1+2\rho V_1^\prime),\\
\nonumber
B &=& h^2+g_1V_1+g_2(V_1+2\rho V_1^\prime),\\
\nonumber
C &=& g_1 g_2,\\
\nonumber
D &=& \sqrt{B^2-4AC},\\
E &=& g_1+g_2.
\end{eqnarray}
In our simple truncation with $S=Z_1+Z_1^\prime\,\rho_0$, $V=V_1$, $Z_2^\prime=V_2=V_3=V_1^\prime=V_2^\prime=V_3^\prime=0$, and at $\sigma=\sigma_0$, the integrand in eq. \eqref{eqFlowEffectivPotentialSystematic} becomes mostly independent of the spatial momentum. The integral can than be carried out and we find
\begin{eqnarray}
\nonumber
&& \zeta(\rho,\sigma_0)= (1-\frac{\eta}{d+2}) \frac{\sqrt{2}\,v_d}{d \,D}\\
\nonumber
&& \Bigg{(}\left(\sqrt{B+D}\frac{E}{\sqrt{C}}-2\sqrt{B-D}\right)\,\text{coth}(\frac{\sqrt{B-D}}{\sqrt{8A}\,T})\\
\nonumber
&& + \left(2\sqrt{B+D}-\sqrt{B-D}\frac{E}{\sqrt{C}}\right)\,\text{coth}(\frac{\sqrt{B+D}}{\sqrt{8 A}\,T})\Bigg{)},\\
\end{eqnarray}
with
\begin{eqnarray}
\nonumber
A & = & V^2,\\
\nonumber
B &=& S^2+2 V(k^2+U^\prime+\rho U^{\prime\prime}),\\
\nonumber
C &=& (k^2+U^\prime+2\rho U^{\prime\prime})(k^2+U^\prime),\\
\nonumber
D &=& \sqrt{B^2-4AC},\\
E &=& 2(k^2+U^\prime+\rho U^{\prime\prime}).
\end{eqnarray}
That the momentum integral can be performed analytically is a nice feature of the cutoff \eqref{eqCutoff}. The limit $T\rightarrow0$ is obtained by substituting the $\text{coth}$ functions with unity. 

The flow of the effective potential contains a subtlety that can be seen in the limit $V_i\rightarrow 0$ ($i=1,2,3$), where we find
\begin{eqnarray}
\nonumber
\partial_t U(\rho,\sigma)&=& \eta \rho U^\prime+2 v_d\int_0^{\sqrt{k^2-m^2}}dp\,p^{d-1}\\
\nonumber
&& (2k^2-\eta(k^2-p^2-m^2)+\partial_tm^2)\\
&& \left(\frac{g_1+g_2}{2h\sqrt{g_1 g_2}}\,\text{coth}(\frac{\sqrt{g_1 g_2}}{2\sqrt{h}\,T})+\frac{1}{h}\right).
\end{eqnarray}
The term $1/h$ in the last line is not present if $V_1$ is set to zero from the outset. If $Z_1$ is independent of $\rho$, this term is independent of $\rho$ and gives only an overall shift of the effective potential.

\section{Flow equations for kinetic coefficients}
\label{sectFlowofkineticcoefficients}
We show in this appendix our results for the flow equation of the kinetic coefficients $S$, $\bar{A}$ and $V$. We neglect all contributions from momentum dependent vertices. In other words, we use $\rho$-independent constants $S=Z_1+\rho_0Z_1^\prime$, $\bar{A}=\bar{Z}_2$ and $V=V_1$. In our truncation with $Z_2^\prime=V_2=V_3=0$, and with the cutoff \eqref{eqCutoff}, we can perform all momentum integrations analytically, leading us to
\begin{widetext}
\begin{eqnarray}
\nonumber
\partial_t V&=&\eta V-(1-\frac{\eta}{d+2})\,T\sum_n\\
\nonumber
&&\frac{32\,{v_d}\,k^{2 + d}{\lambda }^2{{\rho }_0}
    \left( k^2\left( S^2 + k^2V \right)  + 
      \left( S^2 + 2k^2V \right) \lambda {{\rho }_0} - 
      2V\left( S^2 + k^2V + V\lambda {{\rho }_0} \right) {{{\omega }_n}}^2 - 
      3V^3{{{\omega }_n}}^4 \right) }{d
    {\left( k^4 + 2k^2\lambda {{\rho }_0} + 
        \left( S^2 + 2k^2V + 2V\lambda {{\rho }_0} \right) 
         {{{\omega }_n}}^2 + V^2{{{\omega }_n}}^4 \right) }^3},\,\\
         \nonumber
\partial_t S&=&\eta S-(1-\frac{\eta}{d+2})\,T\sum_n\\
\nonumber
&&\frac{32\,{v_d}\,k^{2 + d}S{\lambda }^2{{\rho }_0}
    \left( k^4 - 2\lambda {{\rho }_0}\left( k^2 + \lambda {{\rho }_0} \right)  + 
      S^2{{{\omega }_n}}^2 + 2V\left( k^2 - \lambda {{\rho }_0} \right) 
       {{{\omega }_n}}^2 + V^2{{{\omega }_n}}^4 \right) }{d
    {\left( k^4 + 2k^2\lambda {{\rho }_0} + 
        \left( S^2 + 2k^2V + 2V\lambda {{\rho }_0} \right) 
         {{{\omega }_n}}^2 + V^2{{{\omega }_n}}^4 \right) }^3},\,\\
\frac{\partial_t \bar{A}}{\bar{A}} &=&-\eta=-\,T\sum_n\frac{16\,{v_d}\,k^{2 + d}{\lambda }^2{{\rho }_0}}
  {d{\left( k^4 + 2k^2\lambda {{\rho }_0} + 
        \left( S^2 + 2k^2V + 2V\lambda {{\rho }_0} \right) 
         {{{\omega }_n}}^2 + V^2{{{\omega }_n}}^4 \right) }^2}.\,\,
\label{eqflowofVSEta}
\end{eqnarray}
\end{widetext}
Here, $d$ is the number of spatial dimensions, $v_d$ and $\omega_n$ are as in \eqref{eqabbeffpot}. The Matsubara sums over $n$ can be performed analytically again by using \eqref{eqMatsubara} and derivatives thereof.

In the limit $T\rightarrow0$, the Matsubara frequencies are continuous $\omega_n\rightarrow q_0$ and the sum becomes an integral $T\sum_n\rightarrow \frac{1}{2\pi}\int_{-\infty}^\infty d q_0$. The expressions for $\eta$ and $\partial_t S$ in eq. \eqref{eqflowofVSEta} agree with those derived in \cite{Wetterich:2007ba}, while our result for $\partial_t V$ corrects an error in a first version of \cite{Wetterich:2007ba} (in equation (C.8)).

\section{Propagator and dispersion relation}
\label{sectPropDisp}
The inverse propagator is given by the second functional derivative of the effective action
\begin{eqnarray}
\nonumber
\Gamma^{(2)}&=&\begin{pmatrix} \overset{\rightharpoonup}{\delta}_{\phi_1(-q)} \\ \overset{\rightharpoonup}{\delta}_{\phi_2(-q)} \end{pmatrix} \Gamma_k \begin{pmatrix} \overset{\leftharpoonup}{\delta}_{\phi_1(p)}, & \overset{\leftharpoonup}{\delta}_{\phi_2(p)} \end{pmatrix} \\
&=& G^{-1} \delta(p-q),
\end{eqnarray}
and we find from the truncation \eqref{derivativeexpansion}
\begin{eqnarray}
G^{-1}&=&\begin{pmatrix} H+2J+(V_1+2\rho V_1^\prime)q_0^2 &\hspace{-0.2cm},& -q_0 \sqrt{2K} \\ q_0 \sqrt{2K}&\hspace{-0.2cm},& H+V_1 q_0^2 \end{pmatrix}.\quad
\end{eqnarray}
Here we use the abbreviations
\begin{eqnarray}
\nonumber
H&=&Z_2\vec{p}^2-V_3\vec{p}^4+U^\prime\\
\nonumber
J&=&\rho Z_2^\prime\vec{p}^2-\rho V_3^\prime \vec{p}^4+\rho U^{\prime\prime}\\
2K&=&\left( Z_1+\rho Z_1^\prime-2(V_2+\rho V_2^\prime)\vec{p}^2\right)^2.
\end{eqnarray}
At zero temperature, we can analytically continue to real time $q_0\rightarrow i\omega$, and find
\begin{eqnarray}
G^{-1}&=&\begin{pmatrix} H+2J-(V_1+2\rho V_1^\prime)\omega^2 &\hspace{-0.2cm},& -i\omega \sqrt{2K} \\ i\omega \sqrt{2K} &\hspace{-0.2cm},& H-V_1\omega^2 \end{pmatrix}.\quad
\end{eqnarray}

The dispersion relation is found from the on shell condition
\begin{equation}
\text{det}\,G^{-1}=0
\end{equation}
which yields
\begin{eqnarray}
\nonumber
&& H^2+2HJ-2\left(H(V_1+\rho V_1^\prime)+J V_1+K)\right)\,\omega^2\\
&& +\,V_1(V_1+2\rho V_1^\prime)\,\omega^4=0.
\end{eqnarray}
The solutions for $\omega$ define the dispersion relation. We find two branches, according to
\begin{eqnarray}
\nonumber
(\omega_\pm^2) &=& \frac{1}{V_1(V_1+2\rho V_1^\prime)}\Bigg{(}H(V_1+\rho V_1^\prime)+J V_1+K\\
\nonumber
&& \pm\bigg{(}(K+J V_1)^2+2H\left(K(V_1+\rho V_1^\prime)-J V \rho V_1^\prime\right)\\
&&+H^2(\rho V_1^\prime)^2\bigg{)}^{1/2}\Bigg{)}.
\end{eqnarray}
In the phase with spontaneous symmetry breaking, the $(+)$ branch of this solution is an "optical mode", while the $(-)$ branch is a sound mode. The microscopic sound velocity is $c_S=\frac{\partial \omega}{\partial p}\big{|}_{p=0}$. Using $\rho=\rho_0$, $U^\prime=0$, $U^{\prime\prime}=\lambda$ and $Z_2=1$, we find
\begin{equation}
c_S^2=\frac{1}{\frac{(Z_1+\rho_0 Z_1^\prime)^2}{2\lambda\rho_0}+V_1}=\frac{2\lambda \rho_0}{S^2+2\lambda \rho_0 V}.
\end{equation}
The "optical mode" has at vanishing spatial momentum the frequency
\begin{equation}
\omega_{+}^2(\vec{q}^2=0)=\frac{2\lambda \rho_0}{V_1+2\rho_0V_1^\prime}+\frac{(Z_1+\rho_0Z_1^\prime)}{V_1(V_1+2\rho_0V_1^\prime)}
\end{equation}
which diverges $\omega_{+}^2\rightarrow\infty$ in the limit $V_1\rightarrow0$.

\end{appendix}

\end{document}